%% file: snehurka.tex
\documentclass{aa}
\usepackage{amsmath,graphicx,amssymb}
\usepackage[varg]{txfonts}
\usepackage{natbib}
\usepackage{pst-eps}
\usepackage{subfloat}
\usepackage{stfloats}
\bibpunct{(}{)}{,}{a}{}{,}

\newcommand{\zav}[1]{\left(#1\right)}
\newcommand{\hzav}[1]{\left[#1\right]}
\newcommand{\szav}[1]{\left\{#1\right\}}
\newlength\staretab
\newcommand{\Teff}{\mbox{$T_\mathrm{eff}$}}
\newcommand\de{\text{d}}
\newcommand\x[1]{\ensuremath{#1_\text{X}}}
\newcommand\lx{\ensuremath{\x L}}
\newcommand\vnek{\ensuremath{\varv_\infty}}
\newcommand\msr{\ensuremath{M_\odot\,\text{yr}^{-1}}}
\newcommand\kms{\ensuremath{\text{km}\,\text{s}^{-1}}}
\newcommand\ers{\ensuremath{\text{erg}\,\text{s}^{-1}}}
\newcommand\diag{$T_\text{eff}$ vs.~$\log g$}

\begin{document}

\title{Stellar wind models of subluminous hot stars}

\author{J.~Krti\v{c}ka\inst{1} \and J. Kub\'at\inst{2} \and
I.~Krti\v{c}kov\'a\inst{1}}


\institute{\'Ustav teoretick\'e fyziky a astrofyziky, Masarykova univerzita,
           Kotl\'a\v rsk\' a 2, CZ-611\,37 Brno, Czech
           Republic
           \and
           Astronomick\'y \'ustav, Akademie v\v{e}d \v{C}esk\'e
           republiky, Fri\v{c}ova 298, CZ-251 65 Ond\v{r}ejov, Czech Republic}

\date{Received}

\abstract{Mass-loss rate is one of the most important stellar parameters. Mass loss via stellar winds may influence stellar evolution and modifies stellar
spectrum. Stellar winds of subluminous hot stars, especially subdwarfs, have not been
studied thoroughly.}
{We aim to provide mass-loss rates as a function of subdwarf parameters and to
apply the formula for individual subdwarfs, to predict the wind terminal
velocities, to estimate the influence of the magnetic field and X-ray ionization on
the stellar wind, and to study the interaction of subdwarf wind with mass loss
from Be and cool companions.}
{We used our kinetic equilibrium (NLTE) wind models with the radiative force 
determined from the radiative transfer equation in the comoving frame (CMF) to
predict the wind structure of subluminous hot stars. Our models solve stationary
hydrodynamical equations, that is the equation of continuity, equation of 
motion, and energy equation and predict basic wind parameters.}
{We predicted the wind mass-loss rate as a function of stellar parameters,
namely the stellar luminosity, effective temperature, and metallicity. The
derived wind parameters (mass-loss rates and terminal velocities) agree with the
values derived from the observations. The radiative force is not able to
accelerate the homogeneous wind for stars with low effective temperatures and
high surface gravities. 
We discussed the properties of winds of individual subdwarfs.
The X-ray irradiation may inhibit the flow in binaries with compact components.
In binaries with Be components, the winds interact with the disk of the Be star.}
{Stellar winds exist in subluminous stars with low gravities or high
effective temperatures. Despite their low mass-loss rates, they are detectable
in the ultraviolet spectrum and cause X-ray emission. Subdwarf stars may lose a
significant part of their mass during the evolution. The angular momentum loss
in magnetic subdwarfs with wind may explain their low rotational velocities.
Stellar winds are especially important in binaries, where they may be accreted
on a compact or cool companion.}


\titlerunning{Stellar wind models of subluminous hot stars}

\authorrunning{J.~Krti\v{c}ka et al.}
\maketitle

\section{Introduction}

Mass loss via stellar winds may influence the evolution of stars and determine
their interaction with interstellar environment. Stellar wind also modifies the
emergent spectrum and is therefore  important for the diagnostics of stars.

Radiatively driven stellar winds exist in many types of hot stars \citep[for
a review]{pulvina} particularly in hot subluminous stars. The subluminous
stars are in the late phases of their evolution and their luminosities are lower
than those of corresponding main sequence stars \citep[e.g.,][]{sam1}. Hot
subdwarfs are typical subluminous stars, which consist of a bare helium burning
stellar core stripped of its envelope during the previous evolution
\citep{durman}.

It is not clear how a star may end up in such an evolutionary phase.
There are more possible evolutionary channels that lead to different types of
subluminous objects. Helium low-luminosity stars may originate as a merger of
two white dwarfs \citep{iba,saje,zhaff} or in a late thermal pulse
\citep{icko,argentinci}. Subluminous stars may be also products of red giants,
which were stripped off their envelopes possibly during binary evolution
\citep[e.g.,][]{jinanvoxfordu}.

Hot subdwarfs are frequently members of binaries.
This
may be connected with
their evolutionary state. Subdwarfs are frequently accompanied by various
objects, including white dwarfs, late type stars or substellar objects and, in a
rare cases, Be stars \citep[e.g.,][]{dvoj4,dvoj3}.
  
There is  growing observational interest in winds of subluminous stars.
Ultraviolet (UV) wind line profiles of central stars of planetary nebulae may be
used together with other observables to determine the stellar parameters
\citep{btpau}. Also subdwarf O (sdO) stars show signatures of wind in the
ultraviolet spectral region \citep{jefham}. The X-ray emission of subdwarf stars
is likewise connected with their winds and follows a similar trend as the X-ray
emission of O stars \citep{bufacek}. The wind may be accreted on a compact
companion leading to X-ray sources similar to high-mass X-ray binaries
\citep{dvoj18}. In binaries consisting of a subdwarf star and a compact object,
the missing X-ray emission may provide an upper limit for the wind mass-loss rate
\citep{merne}.

\citet{vinca} predicted the mass-loss rates for hot subdwarfs and discussed
evolutionary and spectroscopic consequences of these winds. However, these models
did not include the hottest subdwarfs and did not predict the terminal
velocities. \citet{un} provides independent predictions of mass-loss rates and
discussed the role of the wind in the radiative diffusion. Although these models
covered a broader range of stellar parameters, the calculations were based on line
force multipliers neglecting, for example, the finite disk factor i.e., they
assumed the star is  a point source of radiation.

The physics of the wind of hot subluminous stars is generally complex. These
winds are, to some extent, similar to the stellar wind of main-sequence B~stars,
where the effects of multicomponent flow and inefficient shock cooling may be
important \citep{krtzatmeni,votzameni}. To improve the theoretical description
of stellar winds of subluminous hot stars we here provide their wind models,
predicting the basic wind parameters. Moreover, we study the effects that have not
yet been discussed in the context of stellar winds of subluminous stars. For example,
the X-ray irradiation in binaries with a compact component that may affect the
wind accretion, or the influence of magnetic fields that may lead to rotational
braking. We also discuss the properties of the winds of individual stars that were
not available in the literature.

\section{Description of the CMF wind models}

\begin{table*}
\caption{Adopted stellar parameters of the model grid and predicted wind
parameters for individual metallicities.}
\centering
\label{hvezpar}
\begin{tabular}{ccccrrcccccr}
\hline
\hline
Model&\multicolumn{5}{c}{Stellar parameters}&\multicolumn{2}{c}{$0.1Z_\odot$}&\multicolumn{2}{c}{$Z_\odot$}&\multicolumn{2}{c}{$10Z_\odot$}\\
&$\Teff$ & $R_{*}$ & $M$ & \multicolumn{1}{c}{$L$} &
\multicolumn{1}{c}{$v_\text{esc}$} & $\dot M$ & \vnek & $\dot M$
& \vnek & $\dot M$ & \multicolumn{1}{c}{\vnek} \\
& $[\text{K}]$ & $[{R}_{\odot}]$ & \multicolumn{1}{c}{$[M_\odot]$} & $[L_\odot]$
& \multicolumn{1}{c}{[\kms]}& [\msr] & [\kms] & [\msr] & [\kms] & [\msr] &
\multicolumn{1}{c}{[\kms]} \\
\hline
15-08 & 15000 & 0.8 & 0.5 &   29 &  490& \multicolumn{2}{c}{no wind}    & \multicolumn{2}{c}{no wind}      & $ 2.1 \times 10^{ -14 }$ &  10 \\
15-16 & 15000 & 1.6 & 0.5 &  120 &  340& \multicolumn{2}{c}{no wind}    & $ 2 \times 10^{ -13 }$ &  10     & $ 5.9 \times 10^{ -12 }$ &  1060 \\
15-32 & 15000 & 3.2 & 0.5 &  460 &  240& $ 1.7 \times 10^{ -12 }$ & 230 & $ 7.4 \times 10^{ -11 }$ &  370  & $ 1.5 \times 10^{ -09 }$ &  330 \\
25-02 & 25000 & 0.2 & 0.5 &   14 &  980& \multicolumn{2}{c}{no wind}    & \multicolumn{2}{c}{no wind}      & $ 3.8 \times 10^{ -14 }$ &  30 \\
25-04 & 25000 & 0.4 & 0.5 &   56 &  690& \multicolumn{2}{c}{no wind}    & $ 3.4 \times 10^{ -14 }$ &  20   & $ 1.6 \times 10^{ -11 }$ &  1250 \\
25-08 & 25000 & 0.8 & 0.5 &  220 &  490& $ 4.2 \times 10^{ -12 }$ & 510 & $ 6.2 \times 10^{ -11 }$ &  590  & $ 1 \times 10^{ -10 }$ &  1510 \\
25-16 & 25000 & 1.6 & 0.5 &  900 &  340& $ 1 \times 10^{ -10 }$ & 430   & $ 3.6 \times 10^{ -10 }$ &  730  & $ 5.8 \times 10^{ -10 }$ &  1080 \\
25-32 & 25000 & 3.2 & 0.5 & 3600 &  220& $ 8.4 \times 10^{ -10 }$ & 260 & $ 1.9 \times 10^{ -09 }$ &  360  & $ 6.9 \times 10^{ -09 }$ &  470 \\
35-01 & 35000 & 0.1 & 0.5 &   13 & 1380& \multicolumn{2}{c}{no wind}    & \multicolumn{2}{c}{no wind}      & $ 5.6 \times 10^{ -13 }$ &  2470 \\
35-02 & 35000 & 0.2 & 0.5 &   54 &  980& \multicolumn{2}{c}{no wind}    & $ 2.7 \times 10^{ -14 }$ &  840  & $ 6.1 \times 10^{ -12 }$ &  2560 \\
35-04 & 35000 & 0.4 & 0.5 &  220 &  690& \multicolumn{2}{c}{no wind}    & $ 1.7 \times 10^{ -11 }$ &  1420 & $ 3.4 \times 10^{ -11 }$ &  2290 \\
35-08 & 35000 & 0.8 & 0.5 &  860 &  480& $ 4.1 \times 10^{ -11 }$ & 520 & $ 2 \times 10^{ -10 }$ &  1210   & $ 5.6 \times 10^{ -10 }$ &  1120 \\
35-16 & 35000 & 1.6 & 0.5 & 3400 &  320& $ 7.3 \times 10^{ -10 }$ & 570 & $ 1.8 \times 10^{ -09 }$ &  640  & $ 4.8 \times 10^{ -09 }$ &  880 \\
45-01 & 45000 & 0.1 & 0.5 &   37 & 1380& \multicolumn{2}{c}{no wind}    & $ 5.7 \times 10^{ -14 }$ &  1100 & $ 7.8 \times 10^{ -12 }$ &  1900 \\
45-02 & 45000 & 0.2 & 0.5 &  150 &  970& \multicolumn{2}{c}{no wind}    & $ 3.6 \times 10^{ -12 }$ &  1440 & $ 8.7 \times 10^{ -11 }$ &  2020 \\
45-04 & 45000 & 0.4 & 0.5 &  590 &  680& $ 3.7 \times 10^{ -12 }$ & 620 & $ 2 \times 10^{ -10 }$ &  1170   & $ 6.2 \times 10^{ -10 }$ &  1820 \\
45-08 & 45000 & 0.8 & 0.5 & 2400 &  460& $ 2.9 \times 10^{ -10 }$ & 720 & $ 1.7 \times 10^{ -09 }$ &  1060 & $ 2.1 \times 10^{ -09 }$ &  1930 \\
55-01 & 55000 & 0.1 & 0.5 &   82 & 1380& \multicolumn{2}{c}{no wind}    & $ 1.3 \times 10^{ -12 }$ &  1690 & $ 2.6 \times 10^{ -12 }$ &  5140 \\
55-02 & 55000 & 0.2 & 0.5 &  330 &  970& $ 4.7 \times 10^{ -14 }$ & 500 & $ 6.1 \times 10^{ -11 }$ &  1440 & $ 2.3 \times 10^{ -10 }$ &  2400 \\
55-04 & 55000 & 0.4 & 0.5 & 1300 &  670& $ 7.8 \times 10^{ -11 }$ & 700 & $ 7.1 \times 10^{ -10 }$ &  1400 & $ 1.2 \times 10^{ -09 }$ &  2310 \\
\hline
\end{tabular}
\end{table*}

We used our spherically symmetric stationary wind code \citep{cmf1} for the
calculation of the wind models of subluminous hot stars. The line radiative
force in the models was calculated using the solution of the comoving frame
(CMF) radiative transfer equation with occupation numbers derived from the
kinetic equilibrium (NLTE) equations. For given global stellar parameters, the
model enables us to consistently predict the radial wind structure (i.e. the
radial dependence of density, velocity, and temperature) and to derive the wind
mass-loss rate, $\dot M,$ and terminal velocity, $v_\infty$.

The ionization and excitation state of the wind was calculated from the NLTE
equations. Part of the corresponding models of ions \citep[see][for their
list]{nlteiii} was adopted from the TLUSTY model atmosphere input files
\citep{ostar2003,bstar2006} and part was prepared by us using the Opacity and
Iron Project data \citep{topt,zel0} and data described by \citet{pahole}. In
addition, we included the ions \ion{Na}{vi} and \ion{Mg}{vi}. This was crucial
to get the correct ionization structure and radiative force in the models with
X-ray irradiation. The level populations were used to calculate the line
radiative force from the solution of the CMF radiative transfer equation
\citep{mikuh} and to calculate the radiative cooling and heating \citep{kpp}.
The emergent surface flux (corresponding to the inner boundary condition) was
taken from H-He spherically symmetric NLTE model stellar atmospheres of
\citet[and references therein]{kub}. The hydrodynamical equations (the
continuity equation, equation of motion, and the energy equation) were solved
iteratively together with NLTE and radiative transfer equations to obtain the
radial dependence of level populations, density, radial velocity, and
temperature. 

The line data used for the line-force calculation were extracted from the VALD
database (Piskunov et al. \citeyear{vald1}, Kupka et al. \citeyear{vald2}). We
filled the minor gaps in the line list of lighter elements (with atomic number
$Z\leq20$) using the data available at the Kurucz
website\footnote{http://kurucz.harvard.edu}. We also checked our line list using
the Opacity Project data, concluding that no   significant gaps remain
in our line list.

\begin{figure}[t]
\centering
\resizebox{\hsize}{!}{\includegraphics{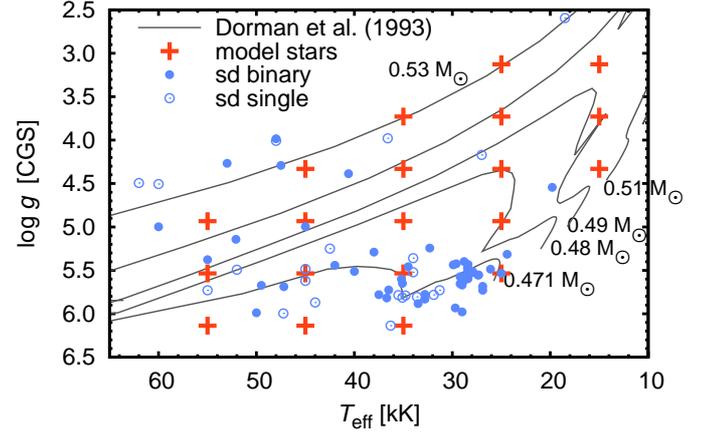}}
\caption{Parameters of studied model stars in \diag\ diagram
(red crosses). Overplotted are solar-metallicity
evolutionary tracks of \citet[][labelled by a
corresponding initial mass]{durman} and the positions of single and binary
member subdwarfs from Tables~\ref{podsam} 
and~\ref{podvoj}.}
\label{sitpotox}
\end{figure}

The adopted parameters of the model stars are given in Table~\ref{hvezpar}
(effective temperature $T_\text{eff}$, radius $R_*$, and mass $M$) together with
the stellar luminosity $L$ and escape speed $v_\text{esc}$. The parameters cover
the evolutionary tracks of horizontal branch stars \citep{durman} and correspond
to the parameters of subdwarfs derived from observations (see
Fig.~\ref{sitpotox}). We assumed a canonical mass $0.5\,M_\odot$ for all models.
The wind models were calculated for three different metallicities (scalling all
elements heavier than helium) $Z=0.1Z_\odot$, $Z=Z_\odot$, and $Z=10Z_\odot$,
where $Z$ is the mass fraction of heavier elements, and $Z_\odot =0.0134 $ is
its solar value. Solar abundances were taken from \citet{asp09}.

\section{Calculated wind models}

\begin{figure}[t]
\centering
\resizebox{\hsize}{!}{\includegraphics{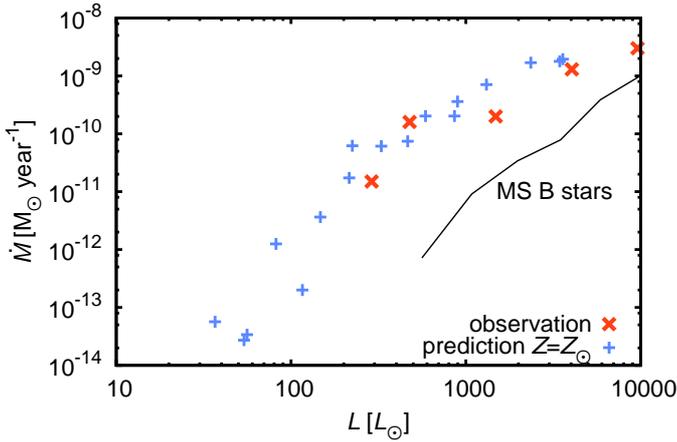}}
\caption{Dependence of the predicted mass-loss rates for $Z=Z_\odot$ (small
blue crosses) in comparison with main sequence mass-loss rates of B stars
\citep[black line,][]{metuje} and observed values for subdwarfs from 
Tables \ref{podsam} and \ref{podvoj} (large red crosses).
The star vZ~1128 is not plotted, since it is a Pop~II star with
different $Z$.}
\label{dmdtl}
\end{figure}

We calculated wind models and predicted the basic wind parameters for adopted
model stars. The resulting mass-loss rates and terminal velocities are given in
Table~\ref{hvezpar} for individual metallicities. The mass-loss rate can be
fitted as
\begin{multline}
\label{dmdtsb}
\log\zav{\frac{\dot M}{1\, \msr }}= -12.61
+\zav{3.78-1.27\log\frac{Z}{Z_\odot}}\log\zav{\frac{L}{10^2L_\odot}}+\\
+\zav{-1.07+0.4\log\frac{Z}{Z_\odot}}\log^2\zav{\frac{L}{10^2L_\odot}}+
1.51\log\frac{Z}{Z_\odot}+\\+
1.09\log\zav{\frac{T_\text{eff}}{10^4\,\text{K}}}.
\end{multline}
As a result of their high effective temperatures, the mass-loss rate of subdwarf
stars is by a factor of about ten higher than the mass-loss rate of main-sequence B
stars with the same luminosities (see Fig.~\ref{dmdtl}). The mass-loss rates
depend strongly on stellar luminosity and on metallicity. The mass loss varies
also with the stellar mass, which is not accounted for in Eq.~\eqref{dmdtsb} owing to
the fixed stellar mass assumed in our models. Based on findings of \cite{vinca},
this dependence is expected to be relatively weak.

\begin{figure}[t]
\centering
\resizebox{\hsize}{!}{\includegraphics{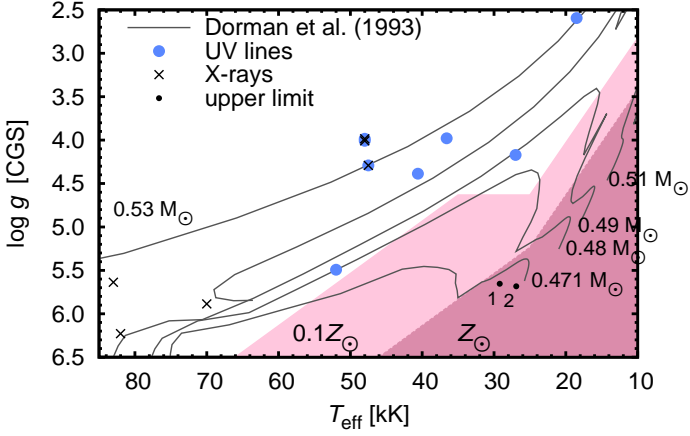}}
\caption{Position of a region with no predicted wind in the
\diag\ diagram for two different metallicities (two shades of red regions).
Overplotted are the evolutionary tracks of \citet{durman} and the positions of
subdwarf stars with known mass-loss rates derived from
observed UV wind-line profiles (blue circles) and X-ray
emission (black crosses) -- see Tables~\ref{podsam} and \ref{podvoj}.
Small black dots denote the positions of stars CD-30\,11223 (1) and 
PG1232-136 (2).}
\label{sitpotoxhrvit}
\end{figure}

We were unable to calculate converged models for stars with high surface
gravities and low effective temperatures. The effect is stronger at low
metallicities. These models are denoted as "no wind" in Table~\ref{hvezpar}. The
failed convergence of the models indicates that the radiative force is too weak
to drive a wind. To test this, we calculated additional models with a fixed
hydrodynamical structure, where we compared the radiative force with the gravity
for four different fixed mass-loss rates equal to $10^{-12}\,\msr$,
$10^{-13}\,\msr$, and $10^{-14}\,\msr$ \citep[for details see][]{metuje}. In all
these models, the radiative force was lower than the gravity force. This result
supports the conclusion that the radiative force is not able to drive a
homogeneous wind and that there is no (hydrogen or helium dominated) wind in the
mentioned cases. The position of the stars with no wind is depicted in
Fig.~\ref{sitpotoxhrvit} by the red  shaded area.

\begin{figure}[t]
\centering
\resizebox{\hsize}{!}{\includegraphics{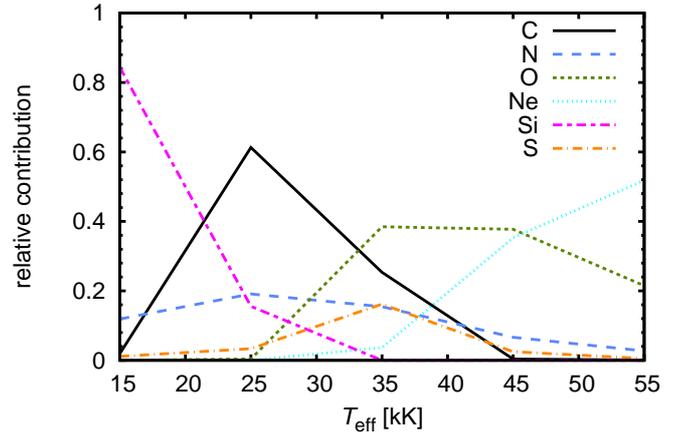}}
\caption{Relative contribution of individual elements to the radiative force
at the critical point of the models with $Z=Z_\odot$ as a function of effective
temperature. Here we plot the results from  models 15-16, 25-08, 35-04, 45-04,
and 55-04.}
\label{grelz1}
\end{figure}

The winds are driven mainly by the heavier element lines of C, N, O, Ne, and Si (see
Fig.~\ref{grelz1}). The contribution of individual elements varies depending mainly on
the effective temperature, but also  on the gravity and metallicity. The
element whose dominant ionization stage has resonance lines close to the maximum
of the flux distribution typically contributes to the radiative force most
significantly. The line driving is dominated by \ion{Si}{iv} for coolest stars,
while more numerous lines of \ion{C}{iii} become more significant in hotter
stars. The flux with energies higher than that of the Lyman jump becomes
significant for line driving in stars with $T_\text{eff}\approx35\,000\,\text{K}$
and, consequently, \ion{O}{iv} is important for the wind driving, while numerous
lines of \ion{Ne}{v} mostly drive the wind in the hottest stars. The plentiful
iron lines that are the most efficient wind driver in O stars do not strongly
contribute to the radiative force in subdwarfs \citep{vinca}. Contrary to less
numerous but stronger lines of lighter elements, the iron lines remain optically
thin as a result of low iron abundance compared to lighter elements
\citep[e.g.,][]{pusle,vikolamet,metuje}. The contribution of iron is only
important at the highest metallicity $Z=10Z_\odot$.

Our models predict a broad range of terminal velocities with the most typical
values of about $300-2000\,\kms$ depending on  stellar parameters (see
Table~\ref{hvezpar}). The ratio of the terminal velocity to the escape speed
$v_\infty/v_\text{esc}$ is typically equal to $1.5$--$2.5$ for solar metallicity
subdwafs, which is slightly lower than $v_\infty/v_\text{esc}=2.6$ found in O
stars \citep{lsl}. The terminal velocity clearly scales with metallicity on
average as $v_\infty\sim Z^{0.2}$. The metallicity dependence of the terminal
velocity and mass-loss rate is stronger than in normal O stars
\citep[c.f.,][]{nlteii}.

We calculated additional models with non-solar helium abundance. Helium neither
significantly contributes to the radiative force nor affects the emergent flux
for subsolar helium abundance. Consequently, our models calculated for
$N(\text{He})/N(\text{H})=0.01$ showed that the subsolar abundance of helium
does not significantly affect the wind mass-loss rate (typically by less than
10\,\%). Our models calculated with enhanced helium abundance (for
$N(\text{He})/N(\text{H})=10$) showed a slightly higher affect on the mass-loss
rate (up to a factor of two) as a result of the influence of helium on the emergent
flux.

Subluminous stars show a very wide range of abundances of individual elements.
Helium may range from hydrogen-dominated atmospheres with subsolar abundance of
helium \citep[e.g.,][]{dvoj47,sam22,sam9,dvoj25} to helium dominated atmospheres
with only small traces of hydrogen \citep{sam1,jefham,dvoj23}. The abundance of
heavier elements, which is crucial for the mass-loss rate determination, shows
comparable variations from star to star \citep{sam15,jefham,dvoj19}. These
abundance differences produce a large diversity of predicted mass-loss rates for
individual stars.

\section{Comparison with observations and with available
theoretical predictions}
\begin{table*}[t]
\caption{Parameters of single subluminous stars.}
\label{podsam}
\begin{tabular}{lccccccccc}
\hline\hline
Star & $T_\text{eff}$ & $M$ & $R_*
$ & observed $\dot M$ & predicted $\dot M$ &
\lx & Source\\
& [K] & $[M_\odot]$ & $[R_\odot]$ & $[\msr]$ & $[\msr]$ & [\ers] \\
\hline
\input{podsam.tex}
\hline
\end{tabular}\\
\tablefoottext{a}{Assumed value.}
\tablefoottext{b}{Member of the globular cluster Messier~3.
Mass-loss rate calculated for $Z=0.1Z_\odot$, which may be more appropriate for
M3.}
\tablebib{\input{ref_podsam.tex}}
\end{table*}
\begin{subtables}
\label{podvoj}
\begin{table*}[t!]
\caption{Parameters of individual subluminous stars in binaries.}
\label{podvoj1}
\begin{tabular}{l@{\hspace{2mm}}cccc@{\hspace{3mm}}c@{\hspace{2mm}}c@{\hspace{2mm}}c@{\hspace{1mm}}c@{\hspace{1mm}}c}
\hline\hline
Star & $T_\text{eff}$ & $M$ & $R_*
$ & $a$ & obs. $\dot M$ & pred. $\dot M$ &
\lx & Comp.\tablefootmark{a} & Source\\
& [K] & $[M_\odot]$ & $[R_\odot]$ & $[R_\odot]$ & $[\msr]$ & $[\msr]$ & [\ers] \\
\hline
\input{podvoj1.tex}
\hline
\end{tabular}
\tablefoottext{a}{Either a spectral type or WD for white dwarf, BD for brown
dwarf, NS for neutron star, and BH for a black hole.}
\tablefoottext{b}{Value during eclipses. Outside eclipses
$\lx=10^{32}\ers$} \tablefoottext{c}{Assumed value.}
\tablefoottext{d}{V2214 Cyg}.
\tablebib{\input{ref_podvoj1.tex}}
\end{table*}
\begin{table*}[t]
\caption{Continuation of Table~\ref{podvoj1}.}
\label{podvoj2}
\centering
\begin{tabular}{lccccccccc}
\hline\hline
Star & $T_\text{eff}$ & $M$ & $R_*
$ & $a$ & obs. $\dot M$ & pred. $\dot M$ &
\lx & Comp. & Source\\
& [K] & $[M_\odot]$ & $[R_\odot]$ & $[R_\odot]$ & $[\msr]$ & $[\msr]$ & [\ers] \\
\hline
\input{podvoj2.tex}
\hline
\end{tabular}
\tablebib{\input{ref_podvoj2.tex}}
\end{table*}
\end{subtables}
We performed a literature search to derive a list of parameters of single
subdwarf stars (see Table~\ref{podsam}) and subdwarf stars in binaries (see
Table~\ref{podvoj}), including their mass-loss rates derived from UV wind line
profiles, the X-ray luminosities \lx\ (preferably in the range 0.2$-$10\,keV, or
upper limits), and orbital separation $a$ for binaries. In Tables~\ref{podsam}
and \ref{podvoj}, we included also the predicted mass-loss rates calculated using
Eq.~\eqref{dmdtsb}, assuming $Z=Z_\odot$. Some stars lie in the region with no
wind in the {\diag} diagram (Fig.~\ref{sitpotoxhrvit}). These stars, for which
we do not predict homogeneous winds, are denoted as ``no wind'' in the Tables.
These stars may still have a pure metallic wind, however, with a very low
mass-loss rate of the order of $10^{-16}\,\msr$ \citep{babelb}.

All listed stars with observed wind line profiles lie outside the region with no
predicted winds in the \diag\ diagram in Fig.~\ref{sitpotoxhrvit}. The X-ray
emission of subdwarf stars is presumably connected with their winds.
Consequently, stars with X-ray emission should also lie outside the region with
no predicted winds. This is also the case for all stars with X-ray emission from
our sample, supporting the reliability of our models. In Fig.~\ref{dmdtl} we
compare the predicted dependence of the mass-loss rate on luminosity with
available values derived from observations. This comparison shows that the
results of our models are also reliable  quantitatively.

In binaries consisting of the subdwarf and a compact companion, the X-ray
emission may originate from the accretion of the wind on the companion.
Consequently, their X-ray luminosity is proportional to the wind mass-loss rate.
The upper limit of $\lx<1.5\times10^{29}\,\text{erg}\,\text{s}^{-1}$ in
CD-30\,11223 thus provides an upper limit for the mass-loss rate $\dot
M<3\times10^{-13}\,\text{M}_\odot\,\text{yr}^{-1}$ \citep{merne}. The estimate
of the upper limit of the mass-loss rate $\dot M\leq10^{-13}\,\msr$ in
PG1232-136 \citep{merne} is poorly constrained as a result of unknown efficiency
for the conversion of accretion power to X-ray luminosity. In any case, both
CD-30\,11223 and PG1232-136 lie in the region with no winds in
Fig.~\ref{sitpotoxhrvit}. Since only the upper limits of their mass-loss rates are
available, this is consistent with our models.

\begin{figure}[t]
\centering
\resizebox{\hsize}{!}{\includegraphics{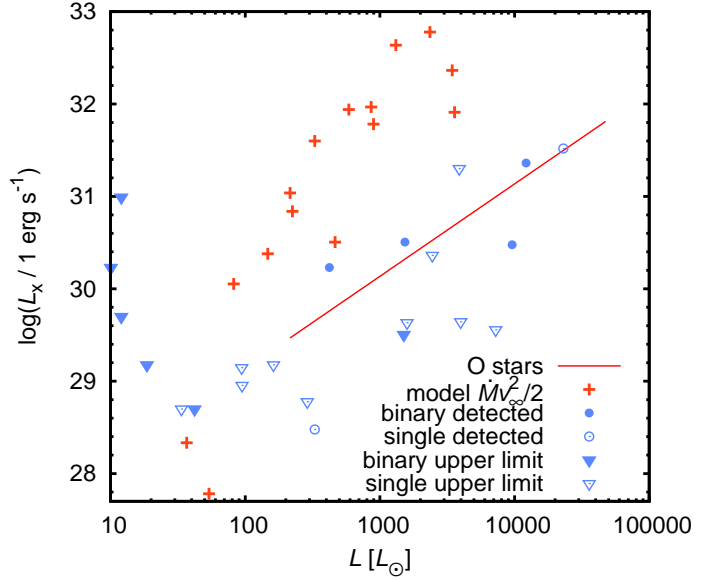}}
\caption{Dependence of X-ray luminosity on the stellar luminosity for
subdwarf stars. Individual symbols refer to the X-ray
detected binary (filled blue circles) and single (empty blue
circles) subdwarf stars and available upper X-ray detection
limit in binary (filled blue
triangles) and single (empty blue triangles) subdwarf stars from
Tables~\ref{podsam} and \ref{podvoj}. Overplotted is the extrapolation of the
observed mean relation for O stars \citep[solid red line]{naze}
and predicted wind kinetic energy lost per unit of time (for $Z=Z_\odot$,
red plus symbols).
}
\label{lxlbol}
\end{figure}

\begin{figure}[t]
\centering
\resizebox{\hsize}{!}{\includegraphics{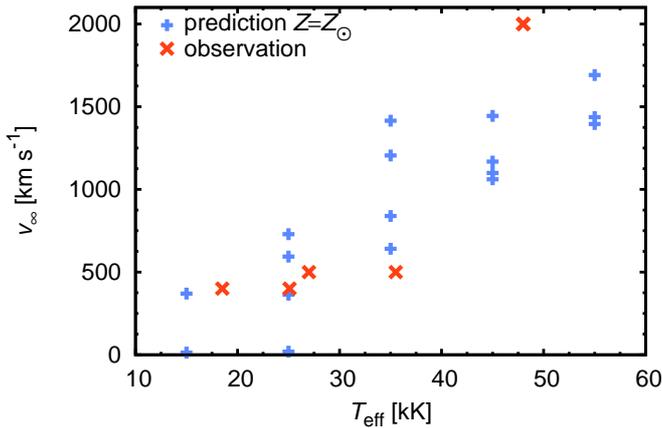}}
\caption{Predicted terminal velocities (blue plus symbols), in comparison
with values derived from observations \citep[red crosses]{jefham}, as a function
of the effective temperature.}
\label{vnekvuni}
\end{figure}
Current models and observations imply that the X-ray emission in O stars
originates in their supersonic winds as a result of different processes
including instabilities in single stars \citep{lusol,ocr,felpulpal}, wind
collision \citep[e.g.,][]{usaci,kapitan,skoda}, and accretion on compact
companion in binaries \citep{davos,laheupet}. The X-ray luminosity in single O
stars and O stars with non-degenerate components is proportional to their stellar
luminosity \citep[$\lx\approx10^{-7}L$, e.g.,][]{igorkar,naze}. The origin of
this observed relationship is not yet fully understood, but it is possibly connected
with the X-ray absorption in the wind \citep{oskal}, the density dependence of
the radiative cooling \citep{simx}, and thin-shell mixing in radiative
wind-shocks \citep{owomix}. The dependence of the X-ray luminosity on the
stellar luminosity in Fig.~\ref{lxlbol} shows that the detected subdwarf stars
follow the extrapolated relation for O stars of \citet{naze}, and also the
available upper X-ray detection limits are, in general, in agreement with this
relation. Moreover, the observed X-ray luminosities are significantly lower than
the wind kinetic energy lost per unit of time $\frac{1}{2}\dot M v_\infty^2$, which indicates that the winds themselves have enough energy to produce the X-rays
(Fig.~\ref{lxlbol}).

In Fig.~\ref{vnekvuni} we compare the predicted terminal velocities with values
derived from observations as a function of the effective temperature. The
terminal velocity is proportional to the escape speed \citep{pulvina}. In our
sample of the stars with winds, the escape speed decreases with effective
temperature (like the gravity, see Fig.~\ref{sitpotoxhrvit}), consequently 
the terminal velocity also decreases with the effective temperature. This roughly
agrees with observations (Fig.~\ref{vnekvuni}) taken from \citet{jefham}.

The predictions of our models agree with the models of \citet{btpau} for central
stars of planetary nebulae \citep{cmf1}. The predicted mass-loss rates are
typically  $20$~\% lower than those derived by \citet{krtzatmeni}. This
difference is caused by the line overlaps that were neglected in  previous
calculations using the Sobolev method. However, our predicted mass-loss rates for
subdwarfs are, on average,  about one magnitude lower than the predictions of
\citet{vinca} and \citet{un}. The models of \citet{vinca} adopt a $\beta$-type
velocity law and also assume the wind terminal velocity to be equal to the
escape speed. Our models consistently calculate the wind velocity from
hydrodynamic equations and typically predict a higher terminal velocity,
$v_\infty/v_\text{esc}\approx1.5$--$2.5$ (see Table~\ref{hvezpar}). This may
significantly affect the predictions, because \citet{vinca} use a global energy
balance to derive the mass-loss rates. Our tests using the predictions of
\citet{vikolamet} showed a factor of 3 difference between the mass-loss rate
predictions of O stars calculated for $v_\infty/v_\text{esc}=1$ and
$v_\infty/v_\text{esc}=2.5$. Solar abundances adopted by \citet{vinca} are also
higher. \citet{un} neglected the finite disk factor (i.e. he replaced the star
with a point source of radiation), which also leads to higher mass-loss rates and
lower terminal velocities \citep{ppk,fa}. Moreover, the line force multipliers
used by \citet{un}, which only approximately describe the wind driving force,
may lead to larger mass-loss rates. The differences are lower for higher
mass-loss rates, which probably reflects a strong decrease of the mass-loss rate close
to the wind limit \citep[c.f.,][]{metuje}.

\section{The winds of single stars}

\subsection{Evolutionary implications}

\citet{vinca} studied the effect of winds on the evolution of subdwarfs and
concluded that line driven winds were not strong enough to significantly modify
evolutionary tracks for horizontal branch stars and to explain the occurrence of
extreme horizontal branch stars in metal-rich clusters. The inspection of the
horizontal branch evolutionary tracks of \citet[see also
\citealt{svitavy}]{durman} shows that the subdwarfs spend most of their time on
a horizontal branch (typically $120-140\,\text{Myr}$) in the region without any
wind (for $Z\lesssim Z_\odot$, see Fig.~\ref{sitpotoxhrvit}). Because stars in
this part of the $T_\text{eff}$ vs.~$\log g$ diagram have typically subsolar
chemical composition, they do not lose any mass during a horizontal branch. Such
stars have higher effective temperatures or lower gravities after leaving the
horizontal branch and they appear in the region with stronger winds in
$T_\text{eff}$ vs.~$\log g$ diagram. However, a typical duration of this type of
evolutionary phase is quite short, typically about $10\,\text{Myr}$
\citep{durman,ostrasit}, consequently the stars lose up to about $0.01\,M_\odot$
of their mass. This is too low to influence the stellar evolution significantly
\citep{vinca}, but may be enough to strip off a thin hydrogen envelope.

However, some of the objects studied here have larger luminosities and mass-loss
rates up to $10^{-9}\,\msr$. Nevertheless, even winds of this greater strength
may affect the stellar evolution only if the corresponding evolutionary states
last for least $100\,\text{Myr}$. This could possibly be  the case of stars with
helium-dominated atmospheres. These stars are typically hotter, consequently they
may lose a more significant fraction of their mass, provided that the
corresponding evolutionary phase is long-lasting.

\subsection{Influence of the magnetic field}

Magnetic field may strongly influence  stellar wind. The effects connected
with the magnetic field include modulation of the X-ray flux in magnetospheres
filled by the wind \citep{donati,malanaze}, trapping of the wind
in corotating magnetosphere \citep{labor,towog}, or the rotational braking
\citep{brzdud}. Although there are positive detections of the magnetic field in
subdwarf stars, a critical assessment of many of them showed negative results
\citep{podland}. On the other hand, the available upper limits of the magnetic
field intensity in subdwarfs still enable a significant influence of the magnetic
field on the wind. This motivates us to study the consequences of magnetic fields in
subdwarfs.

The effect of the stellar wind is characterized by the ratio between magnetic
field energy density and kinetic energy density of the wind, which may be
parameterized by the wind magnetic confinement parameter introduced by
\citet{udo}, \begin{equation} \label{etahve} \eta_*=\frac{B^2R_*^2}{\dot M
v_\infty}, \end{equation} where $B$ is the surface magnetic field strength at
the magnetic equator. The larger the magnetic field energy density is, the
stronger the influence of the magnetic field on the flow is. The magnetic field
significantly affects the wind for $\eta_*\gtrsim1$. It follows from
Table~\ref{hvezpar} that for $Z=Z_\odot$ Eq.~\eqref{etahve} requires a
magnetic field to be as strong as about $100\,\text{G}$ to affect the wind in stars
with the largest mass-loss rates. Such magnetic fields are comparable with the upper
limits of the order of $100\,\text{G}$ given in \citet{podland}. This shows that
magnetic fields might be important for the subdwarf winds, even if no field has
been confirmed yet.

For $\eta_*>1$ the structure of the flow depends on the relation between the
Kepler corotation radius $R_\text{K}=\zav{GM/\Omega^2}^{1/3}$, at which the
centrifugal force acting on a corotating matter balances the gravity, and the
Alfv\'en radius $R_\text{A}/R_*\approx0.29+(\eta_*+0.25)^{1/4}$
\citep[see][]{udorot,malykor}, where the wind kinetic energy density is equal to
the magnetic field energy density. Here $\Omega$ is the stellar angular
frequency of rotation at the stellar surface. The magnetosphere is very dynamic
for slowly rotating stars with $R_\text{A}<R_\text{K}$, whereas circumstellar
clouds may be generated in stars with centrifugal magnetospheres with
$R_\text{A}>R_\text{K}$. Subdwarfs are typically very slowly rotating stars with
$v_\text{rot}\sin i$ of the order of $1\,\text{km}\,\text{s}^{-1}$
\citep{ghrot}, which gives the Kepler corotation radius $R_\text{K}$ of the
order of tens stellar radii. This shows that the case of dynamic magnetospheres
is typical in subdwarf stars.

Magnetic subdwarfs with winds lose angular momentum by magnetized winds. The
rate of this process may be characterized by the spin-down time,
$\tau_\text{spin} = J/\dot J$, where $J$ is the stellar angular momentum. The
spin-down time depends on basic stellar parameters, moment of inertia constant
$k$, polar magnetic field strength $B_\text{p}$, and the wind parameters,
$\tau_\text{spin}\sim kM(v_\infty/\dot M)^{1/2}/(B_\text{p}R_*)$
\citep[Eq.~25]{brzdud}. To estimate the moment of inertia constant $k,$ we used
MESA evolutionary models \citep{mesa1,mesa2}. We started with He core flash
model (with ZAMS mass $1\,M_\odot$ and metallicity $Z=0.02$) and applied large
mass loss that removes the envelope. The models end up with a subdwarf star with
mass $0.46\,M_\odot$. The moment of inertia constant $k=1.2\times10^{-3}$ was
derived from the moment of inertia of the model. 

For stellar parameters corresponding to subdwarfs with the largest mass-loss rates
(e.g., BD+37$^\circ$1977, BD+37$^\circ$442, and HD~49798, see
Tables~\ref{podsam} and \ref{podvoj}), and $B_\text{p}=100\,\text{G}$, we derive
from Eq.~(25) of \citet{brzdud} a spin-down time of the order of 1~Myr. This is
shorter than the evolutionary timescale of subdwarf stars \citep{durman}. The
spin-down time is also significantly shorter than for main-sequence B stars as a
result of a compact core (and low value of $k$, c.f., \citealt{metuje}). The
spin-down time could be even shorter for stars with a stronger field. Moreover,
the magnetic field may be strongly amplified in a merger event \citep{zhumag},
which is one  formation channel of subdwarf stars. We conclude that
magnetic braking can be important in subdwarfs with wind and a magnetic field, and
may possibly explain the low rotational velocities observed in these stars
\citep{ghrot}. A similar effect was proposed by \citet{vinca} for non-magnetic
stars.

\section{The winds of binaries with subdwarf components}

\subsection{Compact companions: the effect of X-ray irradiation}
\label{kapira}

Several subdwarf stars exist in binaries with compact companions.
These binaries are most luminous in the X-ray domain among objects
containing subdwarfs. Their X-ray emission originates in the accretion of wind
by a compact companion. There is a feedback effect of produced X-rays on the
ionization structure of subdwarf stars. We provide wind models that describe
this effect of X-ray irradiation. For our study,we selected model 45-08, which
has parameters close to known subdwarfs in compact binaries (e.g.,
BD+37$^\circ$442, see Table~\ref{podvoj}).

\begin{figure}[t]
\centering
\resizebox{\hsize}{!}{\includegraphics{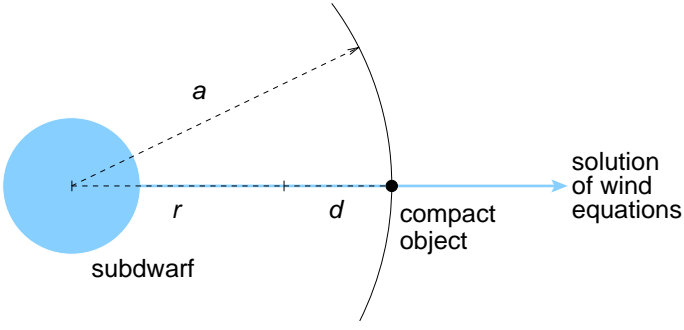}}
\caption{Geometry of the model of a subdwarf wind
irradiated by X-rays from a compact companion.}
\label{trpaslici}
\end{figure}

We studied the effect of X-ray irradiation on the subdwarf wind in a similar way
to our earlier wind-irradiation studies \citep{dvojvit}, i.e., in the
direction to the compact companion (see Fig.~\ref{trpaslici}). The influence of
X-rays is expected to be the strongest in this direction. Moreover, the accreted
wind is assumed to originate from the subdwarf surface that faces the compact
companion. The X-ray source coincides with the compact companion in our models.
The influence of the X-ray irradiation is taken into account as an additional
term in the mean specific intensity,
\begin{equation}
\label{xneutron}
J_\nu^\text{X}=\frac{L_\nu^\text{X}}{16\pi^2d^2}\text{e}^{-\tau_\nu(r)},
\end{equation}
where $L_\nu^\text{X}$ is the luminosity per unit of frequency (we assume
$L_\nu^\text{X}\sim\nu^{-1}$), $d=|a-r|$ is the distance to the X-ray
source from the point in the wind with radius $r$
(neglecting the radius of a compact companion),
the optical depth between a given point and the X-ray source is
\begin{equation}
\label{zamlhou}
\tau_\nu(r)=\left|\int_r^a\kappa_\nu(r')\rho(r')\,\de r'\right|,
\end{equation}
and $\kappa_\nu(r')$ is the opacity per unit of mass. In our approach, the X-rays
directly influence only the ionization equilibrium, while other effects are
neglected. We calculate a grid of wind models for different X-ray luminosities
$\x L=\int L_\nu^\text{X}\,\de\nu$ and different binary separations $a$.

The radial velocity may become non-monotonic in the presence of the external
irradiation. In this case, we cannot calculate the CMF line force directly.
However, we calculate the ratio of the CMF and Sobolev line force
\citep[see][]{cmf1} for a model without external X-ray irradiation and use this
ratio to correct the Sobolev line force in the models with external X-ray
irradiation \citep{dvojvit}. We note that, by taking this approach, we neglect non-local
radiative coupling between absorption zones, which occurs in the non-monotonic
winds \citep{rybashumrem,felnik}.

To simplify the calculation of $J_\nu^\text{X}$ in Eq.~\eqref{xneutron}, we use
a density and absorption coefficient in the form of
\begin{equation}
\begin{split}
\rho(r)&=\frac{\dot M}{4\pi r^2 v(r)},\\
\varv(r)&=\min(\tilde \varv(r),\varv_\text{kink}),\\
\kappa_\nu(r)&=\tilde \kappa_\nu^\text{X},
\end{split}
\end{equation}
where $\tilde \kappa_\nu^\text{X}$ is the depth-independent approximation of
X-ray opacity, $\varv_\text{kink}$ is equal to the velocity of the kink,if it
is present in the models, and otherwise $\varv_\text{kink}=\infty$. The fits to
the wind velocity $\tilde \varv(r)$ and absorption coefficient
$\tilde\kappa_\nu^\text{X}$ are derived from the model with no external
irradiation. The wind velocity is fitted as
\begin{multline}
\label{vrfit}
\tilde 
\varv
(r)=\hzav{\varv_1\zav{1-\frac{R_*}{r}}+\varv_2\zav{1-\frac{R_*}{r}}^2}
     \szav{1-\exp\hzav{\gamma\zav{1-\frac{r}{R_*}}^2}},
\end{multline}
where $\varv_1$, $\varv_2$, and $\gamma$ are free parameters of the fit given in
Table~\ref{bilovice}. The polynomial expansion in Eq.~\eqref{vrfit} provides a
better fit of the model wind velocity than a more commonly used $\beta$ velocity
law \citep{betyna}. The X-ray opacity per unit of mass averaged for radii
$1.5\,R_*-5\,R_*$ is approximated as
\begin{equation}
\label{kapafit}
\log\zav{\frac{\tilde \kappa_\nu^\text{X}}{1\,\text{cm}^2\,\text{g}^{-1}}}=
  \left\{\begin{array}{l}
    \min(a_1\log\lambda+b_1,\log a_0)\quad \lambda<\lambda_1,\\
    a_2\log\lambda+b_2,\quad \lambda>\lambda_1,\\
  \end{array}\right.\\
\end{equation}
where $\lambda_1=20.18$. The parameter $\lambda$ is non-dimensional and
has the same value as the wavelength in units of \AA. Here $a_0$, $a_1$, $b_1$,
$a_2$, and $b_2$ are parameters of the fit given in Table~\ref{bilovice}.

\begin{table}
\caption{Parameters of the
opacity and velocity fits, Eqs.~\eqref{vrfit} and \eqref{kapafit},
respectively.}
\centering
\label{bilovice}
\begin{tabular}{lccc}
\hline
\hline
\multicolumn{1}{c}
{Velocity fit}&
$\varv_1$ [$\text{km}\,\text{s}^{-1}$] & $\varv_2$ [$\text{km}\,\text{s}^{-1}$]&
$\gamma$\\
\hline
&1380 & $-$200& $-$190 \\
\end{tabular}

\begin{tabular}{lccccc}
\hline
\multicolumn{1}{c}{Opacity fit}&
$a_0$ & $a_1$ & $b_1$ & $a_2$ & $b_2$ \\
\hline
&115 & 2.121 & $-$0.566 & 1.397 & $-$0.207\\
\hline
\end{tabular}
\end{table}

\begin{figure}[t]
\centering
\resizebox{\hsize}{!}{\includegraphics{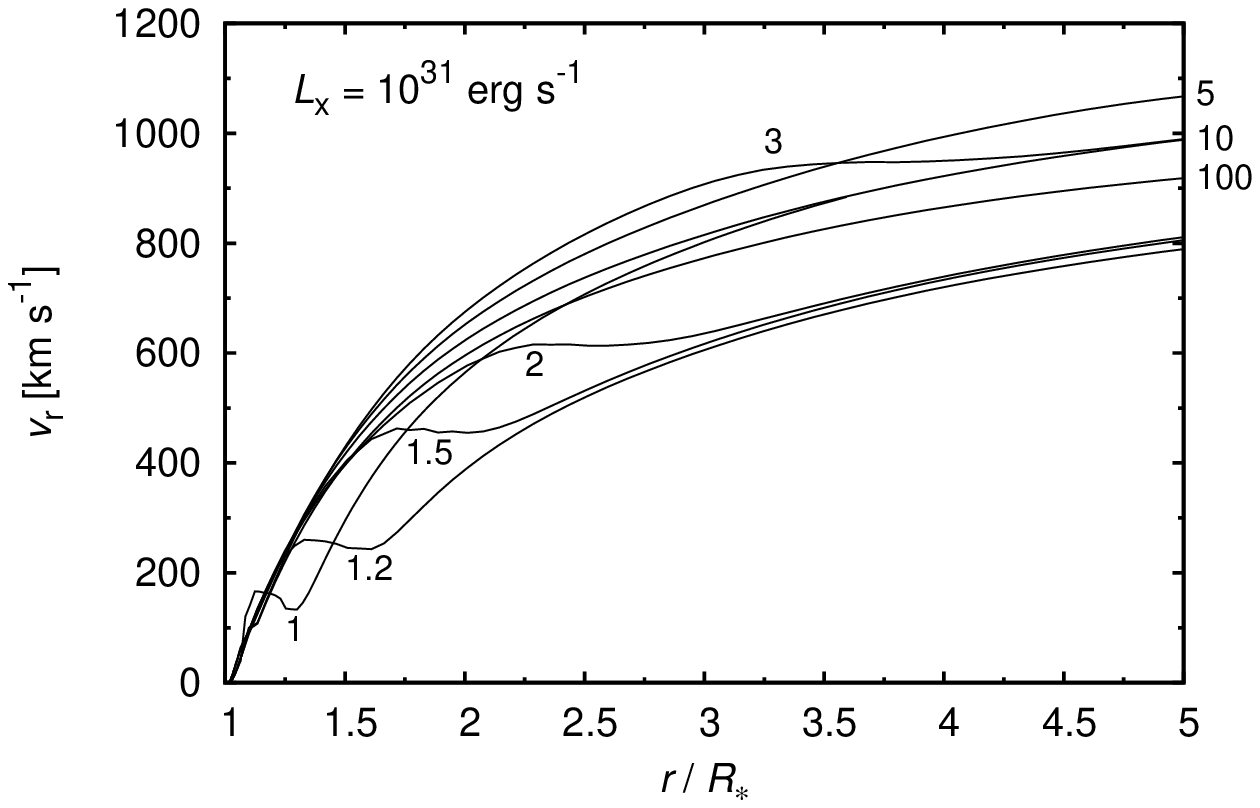}}
\resizebox{\hsize}{!}{\includegraphics{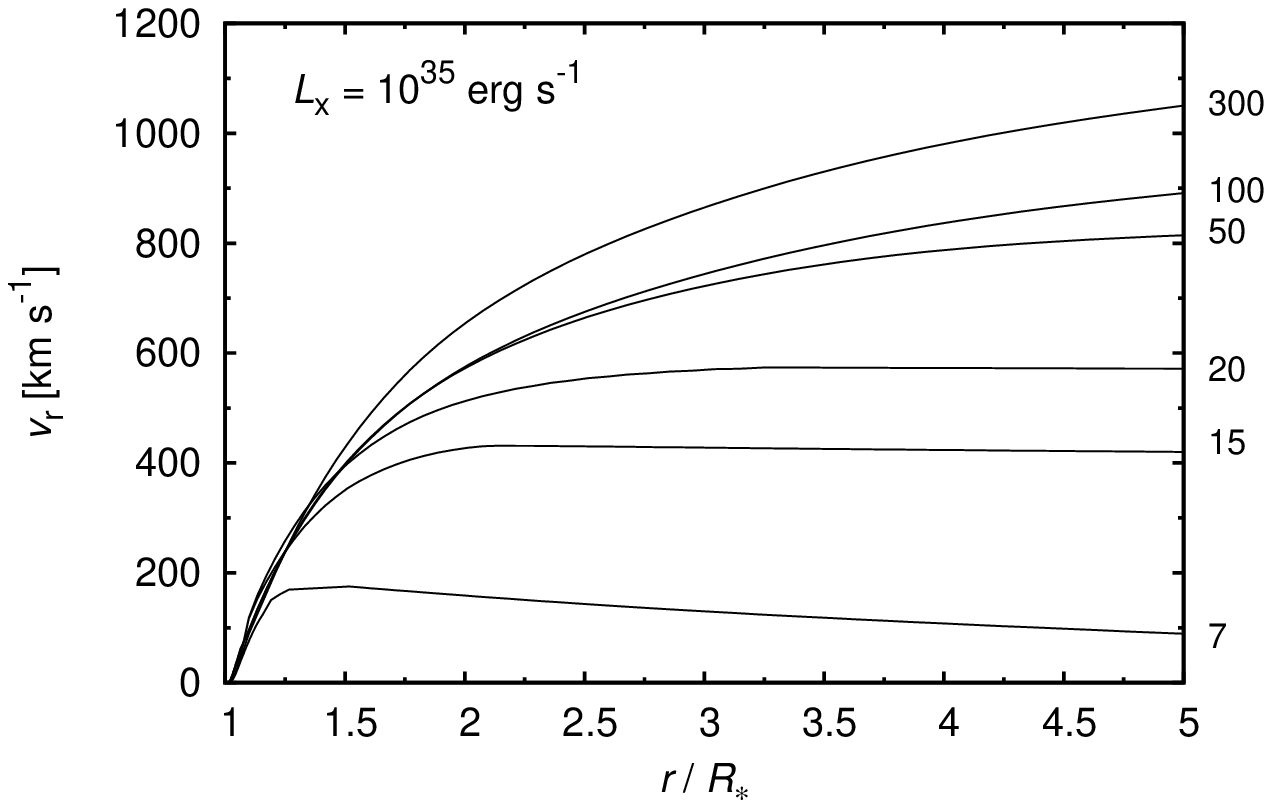}}
\caption{Wind radial velocity as a function of radius in the model 45-08
with external X-ray irradiation (for two different values of
the X-ray luminosity \lx\ given in the plots)
for $Z=Z_\odot$. Individual curves are indicated by the distance
$a$ between the subdwarf and the X-ray source in units of $R_\odot$.}
\label{45-08vr}
\end{figure}

The calculated models for $\lx=10^{31}\,\ers$ and $\lx=10^{35}\,\ers$ are given
in Fig.~\ref{45-08vr} for different distances $a$ between the X-ray source and
the subdwarf. The presence of an X-ray source causes a shift in the wind
ionization towards ions with a higher charge. This affects the radiation force.
For a lower amount of  X-ray irradiation, new ionization states emerge that
are able to drive the wind, and the ionization states with lower charge still remain
 populated. Consequently, the radiative force increases. However, for 
stronger X-ray irradiation, the ions with a lower charge disappear, which leads to the
decrease of the radiative force. This may even lead to wind inhibition in the
direction towards the compact companion \citep{dvojvit}.

This explains the trends given in Fig.~\ref{45-08vr} for models with both
$\lx=10^{31}\,\ers$ and $\lx=10^{35}\,\ers$. For a large distance of the X-ray
source, the wind is not significantly affected by the X-ray irradiation. For
a closer X-ray source, the radiative force increases. Even for a closer X-ray source,
the X-ray ionization is so strong that the wind is not able to accelerate
efficiently and the kink in the velocity law appears close to the position of
the X-ray source \citep{feslop,fero}. If the kink approaches the wind critical
point, where the mass-loss rate of our models is determined, the wind driving is
significantly suppressed, which leads to  wind inhibition. The wind mass flux may
be by one or two orders of magnitude lower in the direction of the compact
companion. However, we do not model this effect in detail, because this would
require time-dependent models.

\begin{figure}[t]
\centering
\resizebox{\hsize}{!}{\includegraphics{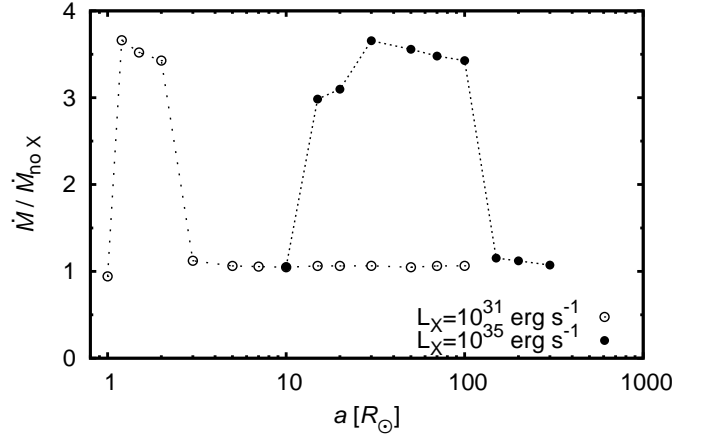}}
\caption{Wind mass-loss rate with X-ray irradiation as a function of the
X-ray source distance. The mass-loss rate is plotted relative to the case
without any X-ray irradiation for two values of X-ray luminosities.}
\label{dmdtx}
\end{figure}

A strong increase of the mass-loss rate for the models with $a\approx2 \,R_\odot
$ in the case of $\lx=10^{31}\,\ers$ and $a\approx50 \,R_\odot $ for
$\lx=10^{35}\,\ers$ (see Fig.~\ref{dmdtx}) resembles the bistability jump in B
supergiants \citep{bista,vikolabis} and has a similar cause. With the increasing
influence of X-rays, the radiative force slightly increases in such a way that
the density rises and the wind ionization shifts towards ions with a lower
charge. The most important is a change of a dominant ionization state of iron
from \ion{Fe}{vi} to \ion{Fe}{v}, but other elements play a role as well. The
ionization shift leads to an increase of the radiative force and mass-loss rate.
This effect is likely limited to very specific stellar parameters (effective
temperature, metallicity) and may not be common among subdwarfs. The range of
X-ray source distance $a$ for which the mass-loss rate significantly increases
depends on the X-ray luminosity. However, if the curves are plotted as a
function of $a-R_*$, they are similar for different X-ray luminosities.

\begin{figure}[t]
\centering
\resizebox{\hsize}{!}{\includegraphics{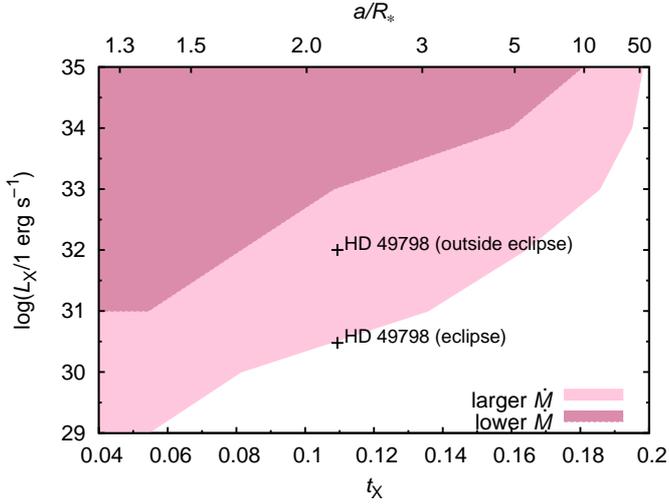}}
\caption{Diagram of
the X-ray luminosity \lx\ versus the optical depth parameter
\x t. The dark red area indicates the region of
parameters, where X-rays inhibit the stellar wind, the light red area indicates
the region of parameters, where X-rays
increase the mass-loss rate, and there is
no strong effect of X-rays on the mass-loss rate in the white area.
Overplotted is the position of the star HD~49798, corresponding to time period of
X-ray eclipse and time period outside the X-ray eclipse \citep[with
$\lx\approx10^{32}\,\ers$,][]{dvoj18}.}
\label{45-08tx}
\end{figure}

The derived results can be summarized in the diagram of the X-ray luminosity
versus the optical depth parameter 
\citep[introduced by][]{dvojvit}:\begin{equation}
\label{tx}
\x t=\frac{\dot M}{\varv_\infty}\zav{\frac{1}{R_*}-\frac{1}{ a }}
\zav{\frac{10^3\,\text{km}\,\text{s}^{-1}\,1\,R_\odot}
{10^{-8}\,{M}_\odot\,\text{year}^{-1}}}.
\end{equation}
The optical depth parameter is proportional to the optical depth between the
X-ray source and the critical point of the stellar wind \citep{dvojvit}. For low
X-ray luminosities (or large \x t, see Fig.~\ref{45-08tx}) the influence of
X-rays can be neglected. For higher X-ray luminosities or slightly lower \x t\
the X-ray irradiation strongly affects the wind ionization state, which in a
particular case leads to the increase of the wind-mass flux in a direction
towards the companion. For even higher X-ray luminosities (lower \x t), the
X-rays start to disrupt the wind and decrease the wind mass flux. The position
of HD~49798 close to the border of the area with lower mass-loss rate indicates
that the X-ray emission of this star may be self-regulated, similar to some
high-mass X-ray binaries \citep{dvojvit}. The self-regulated X-ray emission
means that a higher X-ray emission may lead to the wind inhibition and therefore
to the decrease of \lx, whereas a lower X-ray emission leads to an increase of
the mass flux and increase of \lx. Analogous effects were also predicted  in
main-sequence star winds \citep{pars,dvojvit}.

Our models provide just the wind structure in the direction of the compact
companion, where the effect of X-rays is the strongest. The influence of X-rays
is weaker in other directions forming the photoionization wake \citep{ff}. For a
strong X-ray irradiation, this leads to the dependence of the wind-mass flux on
the location on the stellar surface. The flow is also influenced by the gravity
of the compact object (accretion wake), consequently the numerical simulations
are necessary to study this problem in detail \citep{blondyn,felabon,hacek}.

\subsection{Late-type main-sequence companions: winds in interaction}

Cool main-sequence stars are typical companions of subdwarf stars (see
Table~\ref{podvoj}). These stars may have a solar type (coronal) wind with
typical mass-loss rates $10^{-14}-10^{-11}\,\msr$ and terminal velocities of the
order of $100\,\kms$ that depend on stellar parameters including age
\citep{strelec,nemskot,crasa}. Winds of cool companions may interact with the
wind of a subdwarf star or accrete on the surface of a subdwarf if the subdwarf
wind is either missing or too weak.

The physics of interacting winds is rather complex
\citep[e.g.,][]{skoda,magro,papina}. To understand the basic structure of the
interaction of a cool wind from a hot star and a hot wind from a cool star, we
introduce a simplified picture. We assume that the interaction proceeds through
a series of shocks that form a collisional front. Furthermore, we assume that
the collisional front is globally in equilibrium, which means that the momenta
deposited per unit of time and surface by the wind of both components are equal
in magnitude. Consequently, at the intersection of both components
\citep[e.g.,][]{mihmih,igor},
\begin{equation}
\label{moment}
\rho_\text{sd} \varv_\text{sd}^2=\rho_\text{MS} \varv_\text{MS}^2+\rho_\text{MS}
c_\text{MS}^2,
\end{equation}
where we explicitly included the influence of the cool star wind thermal
energy. Here the subscripts sd and MS denote the wind parameters (density
$\rho$, terminal velocity $\varv$, and thermal speed $c$)
of subdwarf star and cool main-sequence star, respectively.
We assumed that the winds had already reached the corresponding terminal velocities,
and that the subdwarf wind thermal energy can be neglected, since $c_\text{sd}
\ll v_\text{sd}$. Denoting the radial distances of the shock front from the
individual star centres $D_\text{sd}$ and $D_\text{MS}$ and using the continuity
equation, we find
\begin{equation}
\label{momrov}
\frac{D_\text{sd}^2}{D_\text{MS}^2}=\frac{\dot M_\text{sd}\varv_\text{sd}}
{\dot M_\text{MS}\varv_\text{MS}\zav{1+
\dfrac{
c^2_\text{MS}}
{\varv^2_\text{MS}}}},
\end{equation}
with $a=D_\text{sd}+D_\text{MS}$. This equation, in principle, enables us to determine
the fate of both winds, but it cannot be used in practice owing to problems with
the determination of mass-loss rates and velocities of cool star wind.
However, it can at least help us to understand what a typical result
of wind collision could be.

For $D_\text{MS}<R_\text{MS}$, the shock position predicted using
Eq.~\eqref{momrov} would be inside the star. Consequently, the subdwarf wind
accretes directly on the cool companion. Within the Bondi-Hoyle-Lyttleton
accretion model, the wind is accreted from the radius
$r_\mathrm{HL}={2GM_\text{MS}}/{v^2}$, where $v^2=\varv_\text{sd}^2+
\varv_\text{orbit}^2$ ($\varv_\text{orbit}$ is orbital velocity). The
Bondi-Hoyle-Lyttleton accretion radius corresponds to the radius from which the
escape speed is equal to $v$. Because the speed of the subdwarf wind is
typically higher than the escape speed from the main-sequence star, the
accretion radius is lower than the stellar radius $r_\mathrm{HL}<R_\text{MS}$.
Consequently, we can use the radius of the main-sequence star to calculate the
amount of mass accreted per unit of time on the main-sequence star, $\dot
M_\text{acc}\approx\dot M_\text{sd}{R_\text{MS}^2}/{(4a^2)}$, which follows from
geometrical considerations. The mass of cool main-sequence star atmosphere is
from the ATLAS model grid \citep{kurat,casat} of the order of
$10^{-11}-10^{-10}\,M_\odot$. Since the radius of a cool companion is of the
order of $1\,R_\odot$ and the distance is typically also of the same order (see
Table~\ref{podvoj}), the subdwarf star wind replenishes the cool companion
atmosphere within the time period of the order of years. Consequently, the
chemical composition of the cool companion derived from spectroscopy should be
the same as the chemical composition of the subdwarf. This effect was observed
in AA~Dor \citep{vuckodor}, for which we predict the existence of a subdwarf
wind, and which has a sufficiently low semimajor axis (Table~\ref{podvoj}). Given
a typical lifetime of subdwarf stars of the order of $10^8\,\text{years}$
\citep{durman}, in binaries with extreme mass-loss rate of the order of
$10^{-9}\,\msr$, the cool companion may accrete a significant fraction of
the subdwarf's mass. Moreover, the wind is colliding with main-sequence star via a
shock, which creates X-rays with X-ray luminosity $\lx\approx \dot
M_\text{acc}v_\text{sd}^2/2\approx \dot M_\text{sd} v_\text{sd}^2
{R_\text{MS}^2}/(8a^2)$. Because $a$ and $R_\text{MS}$ are of the same order in
many binaries (see Table~\ref{podvoj}), the X-ray source is comparable to the
intrinsic X-ray emission of single subdwarf star (see Fig.~\ref{lxlbol}).
Therefore, the X-ray observations of such objects may show orbital modulation.
Binary HFG~1 is an ideal test case, because it has a relatively close cool
companion and observed X-ray emission (Table~\ref{podvoj}).

Other cases apart from $D_\text{MS}<R_\text{MS}$ are either less common (due to
lower mass-loss rate of a main-sequence star wind) or less observationally
appealing owing to larger orbital separation. For $D_\text{MS}>R_\text{MS}$ and
$D_\text{sd}>R_\text{sd}$, the two winds collide creating an interacting zone
with a complex structure. Such interacting winds are common in hot star wind
binaries \citep[e.g.][]{igor,papina} and may lead to X-ray variability with
orbital period \citep{nazcyg}. For $D_\text{sd}<R_\text{sd}$, the main-sequence
star wind collides with the subdwarf star, which leads to similar effects as for
$D_\text{MS}<R_\text{MS}$, but affects the subdwarf star.

Cool stars are also X-ray sources with typical X-ray luminosities
$10^{26}-10^{30}\,\ers$ \citep{plejadani,norbert,daniel,humburk}. Such X-ray
luminosities typically do not influence the subdwarf wind unless the stars are
very close. For example, for a model 45-08 this means $a\lesssim1.5R_\text{sd}$
for $\lx\gtrsim10^{29}\,\ers$ (see Sect.~\ref{kapira}). These low orbital
separations are not common in subdwarf binaries (see Table~\ref{podvoj}).

\subsection{Be star companions: disk and wind interaction}

In some relatively rare cases, the companion of a subdwarf star is a Be star. Be
stars are fast rotating non-supergiant stars which have, or had, emission lines
owing to an equatorial disk \citep[see][for a recent review]{ricam}. The radiation
from the subdwarf may interact with the disk of the Be star
\citep[e.g.,][]{dvoj4,koukov,dalsikkk}. Moreover, if the subdwarf has a wind
(e.g., in $\varphi$~Per or FY~CMa, see Table~\ref{podvoj}), then there is a
mechanical interaction between a subdwarf wind and a Be-star disk in the region,
where the ram pressures of the Be-star disk and subdwarf wind are equal. Here we
shall derive the location of this interaction region.

Because the wind of the subdwarf star is supersonic,
the mechanical interaction proceeds via a series of shocks (discontinuities, see
\citealt{kurfek}). The location of the interaction region in the equatorial
plane follows from Eq.~\eqref{moment}. The Be-star disks are subsonic to large
distances from the Be star \citep{sapporo,kurfek}, consequently
Eq.~\eqref{moment} can be rewritten as
\begin{equation}
\label{bemoment}
\rho_\text{sd} \varv_\text{sd}^2=\rho_\text{Be} c_\text{Be}^2.
\end{equation}
Inserting the midplane disk density \citep{kom} $\rho_\text{Be}=\dot
M_\text{Be}v_\text{K}(r) /((2\pi)^{3/2}r^2 c_\text{Be} v_r)$, where
$v_\text{K}(r) =\sqrt{GM_\text{Be}/r}$ is the disk orbital (Keplerian) velocity
at radius $r$, and approximating the radial disk velocity $v_r\approx
c_\text{Be} r/R_\text{crit}$ with the disk sonic radius
$R_\text{crit}=\frac{3}{10} \zav{v_\text{K}(R_\text{eqBe})/ c_\text{Be} }^2
R_\text{eqBe}$ \citep{kom}, where for a critically rotating star the equatorial
radius $R_\text{eqBe}=3/2R_\text{Be}$, we derive from Eq.~\eqref{bemoment} the
equation for the distance $r$ of the interaction region from the Be star
\begin{equation}
\label{kubatnemuzeodvodit}
\frac{(a-r)^2R_\text{Be}^{1.5}}{r^{3.5}}=
\frac{10}{3}\sqrt{\frac{\pi} {2}}
\frac{\dot M_\text{sd}}{\dot M_\text{Be}}\frac{
c_\text{Be}^2 \varv_\text{sd}} {v_\text{K}^3 (R_\text{Be})},
\end{equation}
which has to be solved numerically. Here we applied the wind continuity equation
$\dot M_\text{sd}=4\pi(a-r)^2 \rho_\text{sd}\varv_\text{sd}$ valid at the
distance $a-r$ from subdwarf and identity
$v_\text{K}(r)=(R_\text{Be}/r)^{1/2}v_\text{K}(R_\text{Be})$. We note that $a$ is
the distance between centers of both binary components.

For $\varphi$~Per with $M_\text{Be}\approx9\,\text{M}_\odot$ and
$R_\text{Be}\approx5\,R_\odot$ \citep[for spectral type from
Table~\ref{podvoj}]{har}, $ c_\text{Be} \approx20\,\text{km}\,\text{s}^{-1}$,
$\dot M_\text{sd}=3\times10^{-9}\,\msr$,
$\varv_\text{sd}=1400\,\text{km}\,\text{s}^{-1}$, we derive for $\dot
M_\text{Be}=10^{-10}-10^{-8}\,\msr$ \citep{granada} the interaction radius
$r=0.2a-0.6a=50-120\,R_\odot$, which corresponds to the size of a Be-star disk
of $63\,R_\odot$, which was derived from observations \citep{quir,dvoj4}.

\section{Discussion}

\subsection{Inefficient shock radiative cooling}

As a result of the dependence of the radiative force on velocity the wind line
driving leads to an instability, which subsequently steepens in shocks
\citep{luciebila,owor,ocr}. Since the wind density is relatively high in
luminous hot stars, the post-shock material quickly cools down radiatively.
Consequently, the bulk of the wind material has a temperature that is comparable to the
stellar effective temperature \citep{felpulpal}. Because of the dependence of
cooling on the density, the radiative cooling is less effective in weaker winds,
and the wind shocks change from radiative to adiabatic \citep{owomix}. This
effect may explain weak wind line profiles observed in some low-luminosity O
stars \citep{cobecru,nlteiii,lucyjakomy}. In this case, the wind does not cool
effectively after the first shock and stays hot. The shocks appear in the highly
supersonic part of the wind \citep{ocr,felpulpal}, consequently the decreasing
efficiency of the radiative cooling does not affect the mass-loss rate, but may
affect the terminal velocity as a result of the inefficient radiative force.

The problem of inefficient shock radiative cooling has to be addressed with
numerical simulations. \citet{krtzatmeni} provide an analytical estimate for the
ratio of the cooling and hydrodynamic time scales for subdwarf stars. The shock
cooling time $\tau_\text{s}$ can be estimated as a ratio of the kinetic energy
density of post-shock gas and the cooling function,
$\tau_\text{s}\approx3/2(\rho_\text{s}/m_\text{p}\mu)
kT_\text{s}/((\rho_\text{s}/m_\text{p}\mu)^2 \Lambda(T_\text{s}))$, where
$\rho_\text{s}$ is the post-shock density, which can be estimated for infinitely
strong shocks as $\rho_\text{s}=4\rho= \dot M/(\pi r^2v)$, $m_\text{p}$ is the
mass of the proton, $\mu$ is the mean molecular weight, $T_\text{s}$ is the
post-shock temperature, $v$ is radial wind velocity, and
$\Lambda_\text{s}(T_\text{s})$ is the cooling function \citep[see also
\citealt{owomix}]{zelda}. The energy conservation for the post-shock velocity
$v_\text{s}=v/4$ requires that the post-shock temperature is $T_\text{s}=3\mu
m_\text{p} v^2/(16k)$. With the hydrodynamical time-scale $\tau_\text{h}=r/v$
the ratio of the shock cooling time to the hydrodynamic time-scale is
\begin{equation}
\label{zarad}
\frac{\tau_\text{s}}{\tau_\text{h}}=\frac{9}{32}\pi m_\text{p}^2\mu^2\frac{rv^4}
{\Lambda(T_\text{s}) \dot M}.
\end{equation}
The cooling function has a relatively complex dependence on temperature
\citep{rs}, and using calculations of \citet{jiste} for solar chemical
composition
($\Lambda(T_\text{s})\approx
4.4\times10^{-23}(T_\text{s}/10^7\,\text{K})^{1/2}\,\text{erg}\,\text{cm}^3\,
\text{s}^{-1}$, see \citealt{owomix})
the ratio Eq.~\eqref{zarad} can be cast in numerical form: 
\begin{equation}
\label{zaradskal}
\frac{\tau_\text{s}}{\tau_\text{h}}=2\zav{\frac{r}{1\,R_\odot}}
\zav{\frac{v}{10^8\,\kms}}^3\zav{\frac{\dot M}{10^{-9}\,\msr}}^{-1}.
\end{equation}
Eq.~\eqref{zaradskal} shows that for subdwarfs with the mass-loss rate of the
order of $10^{-9}\,\msr$ at a distance of about $1\,R_\odot$, where the wind
velocity is of the order of $100\,\kms$, the shock cooling time is lower than
the hydrodynamic time-scale. Consequently, the shocks are radiative for these
stars and do not significantly alter mean wind structure. For subdwarfs with
lower wind mass-loss rates ($\dot M\lesssim10^{-11}\,\msr$), a significant part
of the wind may not be cooled down efficiently. For these stars, we expect the
same mass-loss rate as predicted by our models, but we expect lower terminal
velocities and weaker wind line profiles.

Besides the radiative cooling,  the adiabatic cooling may also affect the
post-shock temperature \citep{zhek,igor}. Because the typical cooling length owing
to adiabatic cooling is about $r$ \citep[e.g.,][]{owomix}, the adiabatic cooling
time is roughly equal to the hydrodynamical timescale. Consequently,
Eq.~\eqref{zaradskal} also governs  the transition from the radiative to the
adiabatic shocks.

\subsection{Effects connected with multicomponent flow}

The line radiative driving mostly impinges  on heavier elements in hot star winds
(see Fig.~\ref{grelz1}), while the radiative force on hydrogen and helium is
less significant. However, hydrogen and helium constitute the bulk of the wind
material, consequently the momentum has be transferred from heavier ions to
hydrogen and helium  by  Coulomb collisions. In dense winds, this process is
very effective, consequently the winds can be treated as a one-component flow
with equal velocities of all ions \citep{cak76}. However, in low-density
winds the Coulomb collisions become less efficient, which subsequently leads to the
heating of the stellar wind and decoupling of  wind components
\citep{treni,op,nlteii,ufo}.

The multicomponent effects can be important in low-density winds of hot
subdwarfs \citep{krtzatmeni,votzameni}. The importance of the multicomponent
effects can be estimated from the value of the relative velocity difference
$x_{h\text{p}}$ \citep[Eq.~(18) in][]{nlteii} between a given element $h$ and
protons. For $x_{h\text{p}}\lesssim0.1,$ the multicomponent effects are
unimportant, for $x_{h\text{p}}\gtrsim0.1$ the frictional heating typically
influences the wind temperature, and for $x_{h\text{p}}\gtrsim1,$ the wind
components decouple.

\begin{figure}[t]
\centering
\resizebox{\hsize}{!}{\includegraphics{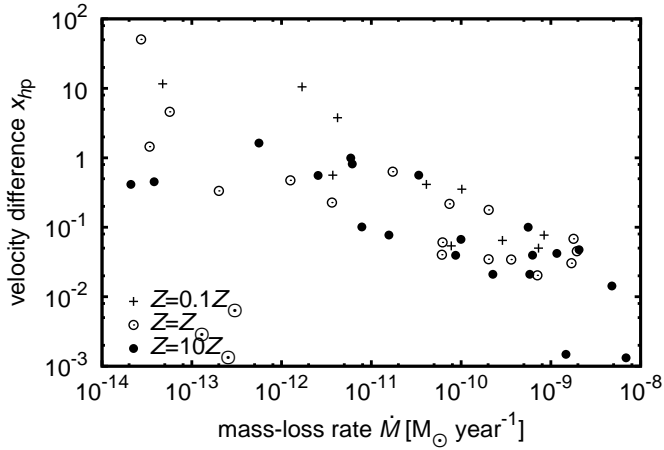}}
\caption{Maximum relative velocity difference between the wind components
and hydrogen \citep[Eq.~(18) in][]{nlteii} as a function
of the wind mass-loss rate for models with different metallicities.}
\label{xhp}
\end{figure}

We calculated the relative velocity difference $x_{h\text{p}}$ according to
\citet{nlteii} for all considered elements in all models. The relative velocity
difference $x_{h\text{p}}$ increases with radius as a result of the decreasing
density and reaches maximum in the outer parts of the wind, where the radiative
acceleration is still significant. We plot the maximum velocity difference
between the heavier ions and hydrogen reached in the wind as a function of
the mass-loss rate in Fig.~\ref{xhp}. The plot is not completely monotonic, because
the value of $x_{h\text{p}}$ depends also on the wind velocity and temperature
(see Eq.~(24) in \citealt{nlteii}), and on the charge and mass of heavier ions
that mostly drive the wind, which also depend on the stellar parameters.

For high wind mass-loss rates $\dot M\gtrsim10^{-10}\,\msr$ the maximum
non-dimensional velocity difference is low, $x_{h\text{p}}<0.1$, and therefore
the flow can be treated as a one-component one. For lower wind mass-loss rates
$10^{-12}\,\msr\lesssim\dot M\lesssim10^{-10}\,\msr$ the maximum non-dimensional
velocity difference is higher, $0.1<x_{h\text{p}}<1$, consequently the wind may
be frictionally heated. This does not affect the mass-loss rate, but may cause
decoupling of the components in the outer wind, because friction decreases with
temperature. For low mass-loss rates $\dot M\lesssim10^{-12}\,\msr$ the wind
decouples. However, only in the star with the lowest mass-loss rates of about
$\dot M\approx10^{-14}\,\msr$ does decoupling occur close to the critical point,
where the wind velocity is equal to the speed of the \citet{abbvln}
radiative-acoustic waves. Because the mass-loss rate of our models is determined
below this region, decoupling mostly affects the wind terminal velocity and not
the mass-loss rate.

Most stars with observed winds have large mass-loss rates $\dot
M\gtrsim10^{-10}\,\msr$ (see Tables~\ref{podsam} and \ref{podvoj}), consequently
the multicomponent effects in their winds are expected to be insignificant. Only
in the case of the star BD~+75$^\circ$325 the mass-loss rate may be so low that the
frictional heating may affect the wind temperature.

\section{Conclusions}

We calculated wind models suitable for subluminous hot stars. Our models derive
level populations from NLTE equations and use hydrodynamical equations with CMF
radiative force to predict the wind structure, i.e., the radial dependence of
density, velocity, and temperature. Our models therefore predict the basic wind
parameters, the mass-loss rate, and the terminal velocity, as a function of
stellar parameters, which include stellar effective temperature, mass, radius,
and chemical composition.

We tested our derived wind parameters, showing that both the predicted mass-loss
rates and terminal velocities agree with values derived from observations. Our
models do not predict any winds for stars with low effective temperatures and
high surface gravities, while the winds may be strong ($\dot
M\approx10^{-10}-10^{-9}\,\msr$) for stars with large luminosities. This result
is in agreement with the position of stars with and without observed X-ray
emission in the $\log g$~vs.~$T_\text{eff}$ diagram. We fitted our derived
mass-loss rates as a function of the stellar luminosity, effective temperature,
and metallicity.

We estimated the impact of the stellar winds on the evolution of subdwarf stars.
Stars with high mass-loss rates $\dot M\gtrsim10^{-9}\,\msr$ may lose a
substantial part of their mass and possibly also angular momentum if they
additionally host a magnetic field. The angular momentum loss via magnetized winds
may explain low rotational velocities observed in some subdwarf stars.

We studied the winds in binaries that host a subdwarf star. The wind radiative
driving in binaries with compact companions may be affected by the presence of a
strong X-ray source. The subdwarf winds are accreted on the surface of nearby
cool main-sequence companions, which affects their apparent chemical composition. In
binaries with a Be-star companion, we predict a mechanical interaction between wind
and disk.

Low-density winds may be affected by the so-called weak wind effect, which is
probably caused by inefficient shock cooling. We provide a formula that
estimates the importance of this effect from basic stellar and wind parameters.
Subdwarf winds are prone to this effect for low mass-loss rates $\dot
M\lesssim10^{-11}\,\msr$. These weak winds may also show multicomponent
structure with the difference between the heavier elements and hydrogen and
helium becoming comparable with the thermal speed. Moderate winds trigger some
abundance peculiarities, while strong winds wash out any peculiarities
\citep{vasam,nedolez,un,vimiri}. Consequently, winds in subluminous stars may be
also important for the explanation of surface abundances.

The stellar winds constitute an important property of luminous subdwarfs. They
may affect their spectra, evolution, and interaction with a potential companion.

\begin{acknowledgements}
We thank Drs.~S.~Mereghetti and P.~N{\'e}meth for  discussing  the topic.
This research was supported by GA\,\v{C}R  13-10589S.
Access to computing and storage facilities owned by parties and projects
contributing to the National Grid Infrastructure MetaCentrum, provided under the
programme Projects of Large Infrastructure for Research, Development, and
Innovations (LM2010005), is greatly appreciated.
\end{acknowledgements}

\end{document}

%% file: podsam.tex
PG0217+155  & 55000 & 0.5 & 0.16 &$   $&$ 2.0\times10^{-11 }$&$   $& 1 \\
LS I +63 198  & 34000 & 0.03 & 0.06 &$   $& no wind &$ <1.4\times10^{29} $& 2, 3 \\
LB 1766  & 36300 & 0.5\tablefootmark{a} & 0.10 &$   $& no wind &$   $& 4, 27 \\
CD-31 4800 (LS VI-03 19) & 44000 & 0.27 & 0.1 &$   $&$ 1.2\times10^{-14 }$&$ <5\times10^{28} $& 3, 5, 6 \\
BD-03$^\circ$2179  & 62000 & 0.21 & 0.43 &$   $&$ 2.8\times10^{-9 }$&$ <2.3\times10^{30} $& 3, 7 \\
BD+75$^\circ$325  & 52000 & 0.5 & 0.21 &$ 1.5\times10^{-11} $&$ 4.8\times10^{-11 }$&$ <6\times10^{28} $& 3, 8 \\
PG0911+456 (DT Lyn) & 31900 & 0.39 & 0.133 &$   $& no wind &$   $& 9 \\
CD-45 5058 (LS 1275) & 75000 & 0.5 & 0.37 &$   $&$ 4.5\times10^{-9 }$&$ <2\times10^{31} $& 2, 3 \\
BD+37$^\circ$1977  & 48000 & 1.8 & 2.2 &$ 6\times10^{-9 }$&$ 1.3\times10^{-9 }$&$ 3.3\times10^{31} $& 3, 10, 11 \\
EC 09582-1137 (V541 Hya) & 34800 & 0.485 & 0.147 &$   $& no wind &$   $& 12 \\
BD+10$^\circ$2179 (DN Leo) & 18500 & 0.55 & 6.2 &$ 1.3\times10^{-9 }$&$ 1.0\times10^{-9 }$&$   $& 10 \\
PG1219+534 (KY UMa) & 33600 & 0.457 & 0.140 &$   $& no wind &$   $& 13 \\
PG1325+101 (QQ Vir) & 35100 & 0.50 & 0.145 &$   $& no wind &$   $& 14 \\
vZ 1128\tablefootmark{b}  & 36600 & 0.5\tablefootmark{a} & 1.2 &$ 1\times10^{-10} $&$ 4.6\times10^{-10 }$&$   $& 31 \\
CD-46 8926 (LSE 153) & 70000 & 0.7 & 0.58 &$   $&$ 4.5\times10^{-9 }$&$ <3.6\times10^{29} $& 3, 15 \\
HD 144941 (CD-26 11229) & 27000 & 0.54 & 1.0 &$ 1.6\times10^{-10} $&$ 8.6\times10^{-11 }$&$   $& 10, 30 \\
LS IV-12.1  & 60000 & 0.16 & 0.37 &$   $&$ 1.7\times10^{-9 }$&$ <4.3\times10^{29} $& 2, 3, 29 \\
HD 149382 (PG1631-039,  BD-03$^\circ$3967) & 35500 & 0.5 & 0.15 &$   $& no wind &$   $& 16, 17 \\
BD+39$^\circ$3226  & 45000 & 0.49 & 0.21 &$   $&$ 7.0\times10^{-12 }$&$ <1.5\times10^{29} $& 3, 18 \\
LSE 263 (CD-51 11879) & 70000 & 0.54 & 0.43 &$   $&$ 4.2\times10^{-9 }$&$ <4.4\times10^{29} $& 2, 3, 15 \\
KPD 1943+4058 (KOI-55) & 28000 & 0.496 & 0.203 &$   $& no wind &$   $& 28 \\
HD 127493 (BD-22$^\circ$3804) & 42500 & 0.21 & 0.18 &$ <2\times10^{-10} $&$ 9.7\times10^{-13 }$&$ <9\times10^{28} $& 3, 5, 32 \\
LS IV+10 9  & 45000 & 0.39 & 0.16 &$   $&$ 1.0\times10^{-12 }$&$ <1.4\times10^{29} $& 3, 19, 29 \\
ALS 11634 (LS IV-14$^\circ$116) & 34000 & 0.485 & 0.2 &$   $&$ 4.5\times10^{-14 }$&$   $& 20, 21 \\
KPD 2109+4401 (V2203 Cyg) & 31300 & 0.5 & 0.16 &$   $& no wind &$   $& 22, 23, 24 \\
BD+28$^\circ$4211  & 82000 & 0.5 & 0.09 &$   $&$ 1.1\times10^{-10 }$&$ 3\times10^{28} $& 3, 25 \\
Feige 110 (GJ 894.3) & 47250 & 0.469 & 0.114 &$   $&$ 1.5\times10^{-13 }$&$   $& 26 \\

%% file: ref_podsam.tex
(1)~\citet{sam1};
(2)~\citet{sam2};
(3)~\citet{bufacek};
(4)~\citet{sam4};
(5)~\citet{sam5};
(6)~\citet{sam6};
(7)~\citet{dvoj26};
(8)~\citet{sam8};
(9)~\citet{sam9};
(10)~\citet{jefham};
(11)~\citet{sam11};
(12)~\citet{sam12};
(13)~\citet{sam13};
(14)~\citet{sam14};
(15)~\citet{sam15};
(16)~\citet{sam16};
(17)~\citet{sam17};
(18)~\citet{sam18};
(19)~\citet{sam19};
(20)~\citet{sam20};
(21)~\citet{sam21};
(22)~\citet{sam22};
(23)~\citet{ostrasit};
(24)~\citet{sam24};
(25)~\citet{sam25};
(26)~\citet{sam26};
(27)~\citet{sam27};
(28)~\citet{sam28};
(29)~\citet{sam29};
(30)~\citet{sam30};
(31)~\citet{sam31};
(32)~\citet{sam32}.

%% file: podvoj1.tex
PG0101+039 (Feige 11) & 27300 & 0.47 & 0.19 & 3.1 &$   $& no wind &$   $& WD & 1, 2 \\
GD 687  & 24400 & 0.47 & 0.25 & 2.3 &$   $& no wind &$   $& WD & 3 \\
$\varphi$ Per (HD 10516,  HR 496) & 53000 & 1.14 & 1.3 & 232 &$   $&$ 2.7\times10^{-9 }$&$   $& B1.5Ve & 4 \\
BD+37$^\circ$442  & 48000 & 0.9 & 1.6 &   &$ 3\times10^{-9 }$&$ 2.4\times10^{-9 }$&$ 2.3\times10^{31} $& WD & 5, 6 \\
CPD -71$^\circ$172  & 55000 & 0.5 & 0.24 &   &$   $&$ 1.8\times10^{-10 }$&$   $& F3-F4IV & 7 \\
HFG 1 (V664 Cas) & 83000 & 0.57 & 0.19 & 3.5 &$   $&$ 2.4\times10^{-9 }$&$ 3.2\times10^{30} $& F-K & 8, 9 \\
KPD 0422+5421 (IQ Cam) & 25000 & 0.36 & 0.17 & 0.93 &$   $& no wind &$   $& WD & 10, 11 \\
V1405 Ori  & 35100 & 0.47 & 0.17 & >2.1 &$   $&$ 1.9\times10^{-14 }$&$   $& F-G & 12 \\
LB 3459 (AA Dor) & 42000 & 0.33 & 0.181 & 1.2153 &$   $&$ 8.3\times10^{-13 }$&$   $& K & 13, 14 \\
Albus 1 (CPD -20$^\circ$1123) & 19800 & 0.49 & 0.62 & >6.5 &$   $&$ 3.9\times10^{-14 }$&$   $&   & 15 \\
HD 49798  & 47500 & 1.5 & 1.45 & 7 &$ 3\times10^{-9 }$&$ 2.7\times10^{-9 }$&$ 3\times10^{30}$\tablefootmark{b} & NS & 16, 17, 18, 69, 72 \\
BD+34$^\circ$1543  & 36700 & 0.47 & 0.14 & 447 &$   $& no wind &$   $& F & 19 \\
HS 0705+6700 (V470 Cam) & 28800 & 0.483 & 0.230 & 0.81 &$   $& no wind &$   $& M & 20 \\
FY CMa (HD 58978) & 45000 & 1.3 & 0.6 & 112 &$   $&$ 1.0\times10^{-9 }$&$   $& B0.5IVe & 21 \\
SDSS J08205+0008  & 26100 & 0.251 & 0.15 & 0.588 &$   $& no wind &$   $& BD & 22 \\
US 708  & 47200 & 0.3 & 0.13 & 1 &$   $&$ 4.4\times10^{-13 }$&$   $& WD & 23 \\
TYC 7709-376-1  & 28400 & 0.461 & 0.179 & 0.963 &$   $& no wind &$   $& dM & 24 \\
EC 10246-2707  & 28900 & 0.45 & 0.17 & 0.84 &$   $& no wind &$   $& dM & 25 \\
Feige 34 (GJ 398.2) & 70000 & 0.55 & 0.14 &   &$   $&$ 1.8\times10^{-10 }$&$ 1.7\times10^{30} $& M & 26, 27 \\
LSS 2018 (KV Vel) & 77000 & 0.63 & 0.157 & 2.1 &$   $&$ 7.5\times10^{-10 }$&$   $& M & 28, 29, 30 \\
Feige 36 (WD 1101+249) & 29700 & 0.45 & 0.12 & 2.4 &$   $& no wind &$ <1.7\times10^{30} $& WD & 65, 66 \\
PG1104+243  & 33500 & 0.47 & 0.13 & 322 &$   $& no wind &$   $& G0 & 31 \\
Feige 48 (KL UMa) & 29900 & 0.46 & 0.215 & 2.1 &$   $& no wind &$   $& K & 32, 33, 71 \\
PG1232-136  & 26900 & 0.45 & 0.16 & >4 &$ <1\times10^{-13} $& no wind &$ <5\times10^{29} $& BH & 42, 65 \\
BD+18$^\circ$2647 (Feige 67) & 60000 & 0.47\tablefootmark{c} & 0.36 &   &$   $&$ 1.6\times10^{-9 }$&$ <3.2\times10^{29} $&   & 27, 34 \\
HW Vir (BD-07$^\circ$3477) & 28500 & 0.48 & 0.197 & 0.853 &$   $& no wind &$   $& dM & 35 \\
CS 1246  & 28450 & 0.39 & 0.19 & 19.6 &$   $& no wind &$   $&   & 37, 38 \\
Feige 80  & 37500 & 0.43 & 0.14 & 530 &$   $&$ 1.1\times10^{-14 }$&$   $& G1V & 26, 39 \\
PG1336-018 (NY Vir) & 32800 & 0.471 & 0.147 & 0.723 &$   $& no wind &$   $& dM & 40, 41 \\
Feige 87  & 27400 & 0.47 & 0.19 & 442 &$   $& no wind &$   $& G4V & 19 \\
CD-30 11223  & 29200 & 0.47 & 0.169 & 0.599 &$ <3\times10^{-13} $& no wind &$ <1.5\times10^{29} $& WD & 42 \\
HD 128220  & 40600 & 0.54 & 0.78 & 520 &$ 2\times10^{-10} $&$ 1.0\times10^{-9 }$&$   $& G0III & 43, 44, 45, 70 \\
PG1432+159  & 26900 & 0.5 & 0.16 &   &$   $& no wind &$ <9.8\times10^{30} $& NS & 66, 67, 68 \\
2M 1533+3759  & 29200 & 0.376 & 0.166 & 0.98 &$   $& no wind &$   $& M5V & 46 \\
PG1605+072 (V338 Ser) & 32300 & 0.5 & 0.28 &   &$   $&$ 3.1\times10^{-13 }$&$   $&   & 47 \\
SDSS J162256.66+473051.1  & 29000 & 0.48 & 0.168 & 0.58 &$   $& no wind &$   $& BD & 48, 49 \\
PG1718+519  & 29000 & 0.5 & 0.12 &   &$   $& no wind &$   $& G4V & 50, 51 \\
MT Ser  & 50000 & 0.6 & 0.13 & 0.9 &$   $&$ 1.2\times10^{-12 }$&$   $& dM & 64 \\
BD+29$^\circ$3070  & 28500 & 0.47 & 0.19 & 586 &$   $& no wind &$   $& F5V & 19 \\
V477 Lyr  & 49500 & 0.508 & 0.172 & 2.21 &$   $&$ 7.4\times10^{-12 }$&$   $& M & 52 \\
KIC 11558725  & 27900 & 0.48 & 0.2 & 20.3 &$   $& no wind &$   $& WD & 53 \\
KPD 1930+2752\tablefootmark{d}  & 35200 & 0.47 & 0.18 & 0.98 &$   $&$ 3.4\times10^{-14 }$&$   $& WD & 54, 55 \\
2M 1938+4603  & 29600 & 0.372 & 0.196 & 0.823 &$   $& no wind &$   $& dM & 56, 57 \\

%% file: ref_podvoj1.tex
(1)~\citet{dvoj1};
(2)~\citet{dvoj2};
(3)~\citet{dvoj3};
(4)~\citet{dvoj4};
(5)~\citet{dvoj5};
(6)~\citet{jefham};
(7)~\citet{dvoj7};
(8)~\citet{dvoj8};
(9)~\citet{dvoj9};
(10)~\citet{dvoj10};
(11)~\citet{dvoj11};
(12)~\citet{dvoj12};
(13)~\citet{dvoj13};
(14)~\citet{dvoj14};
(15)~\citet{dvoj15};
(16)~\citet{dvoj16};
(17)~\citet{dvoj17};
(18)~\citet{dvoj18};
(19)~\citet{dvoj19};
(20)~\citet{dvoj20};
(21)~\citet{dvoj21};
(22)~\citet{dvoj22};
(23)~\citet{dvoj23};
(24)~\citet{dvoj24};
(25)~\citet{dvoj25};
(26)~\citet{dvoj26};
(27)~\citet{bufacek};
(28)~\citet{dvoj28};
(29)~\citet{dvoj29};
(30)~\citet{dvoj30};
(31)~\citet{dvoj31};
(32)~\citet{dvoj32};
(33)~\citet{dvoj33};
(34)~\citet{dvoj34};
(35)~\citet{dvoj35};
(37)~\citet{dvoj37};
(38)~\citet{dvoj38};
(39)~\citet{dvoj39};
(40)~\citet{dvoj40};
(41)~\citet{dvoj41};
(42)~\citet{merne};
(43)~\citet{dvoj43};
(44)~\citet{dvoj44};
(45)~\citet{dvoj45};
(46)~\citet{dvoj46};
(47)~\citet{dvoj47};
(48)~\citet{dvoj48};
(49)~\citet{dvoj49};
(50)~\citet{dvoj50};
(51)~\citet{dvoj51};
(52)~\citet{dvoj52};
(53)~\citet{dvoj53};
(54)~\citet{dvoj54};
(55)~\citet{dvoj55};
(56)~\citet{dvoj56};
(57)~\citet{dvoj57};
(64)~\citet{dvoj64};
(65)~\citet{dvoj65};
(66)~\citet{dvoj66};
(67)~\citet{dvoj67};
(68)~\citet{dvoj68};
(69)~\citet{dvoj69};
(70)~\citet{sam32};
(71)~\citet{dvoj71};
(72)~\citet{dvoj72}.

%% file: podvoj2.tex
KPD 1946+4340  & 34500 & 0.47 & 0.212 & 2.34 &$   $&$ 9.8\times10^{-14 }$&$   $& WD & 58 \\
V2008-1753  & 32800 & 0.47 & 0.138 & 0.56 &$   $& no wind &$   $& BD & 36 \\
NSVS 14256825  & 40000 & 0.419 & 0.188 & 0.80 &$   $&$ 4.9\times10^{-13 }$&$   $& dM & 59 \\
59 Cyg (V832 Cyg, HD 200120) & 52100 & 0.77 & 0.39 & 80 &$   $&$ 7.7\times10^{-10 }$&$   $& B1.5Ve & 60 \\
BD+25$^\circ$4655 (IS Peg) & 38000 & 0.16 & 0.15 &   &$   $&$ 2.8\times10^{-14 }$&$ <0.5\times10^{29} $&   & 27, 61 \\
HS 2231+2441  & 28400 & 0.265 & 0.164 & 1.18 &$   $& no wind &$   $& dM & 62 \\
HS 2333+3927  & 36500 & 0.38 & 0.14 & 1.13 &$   $& no wind &$   $& dM & 63 \\

%% file: ref_podvoj2.tex
(27)~\citet{bufacek};
(36)~\citet{dvoj36};
(58)~\citet{dvoj58};
(59)~\citet{dvoj59};
(60)~\citet{dvoj60};
(61)~\citet{dvoj61};
(62)~\citet{dvoj62};
(63)~\citet{dvoj63}.

%% file: snehurka.bbl
\begin{thebibliography}{}
\bibitem[Abbott(1980)]{abbvln} Abbott, D.~C.\ 1980, \apj, 242, 1183
\bibitem[Af{\c s}ar \& Ibano{\v g}lu(2008)]{dvoj52} Af{\c s}ar, M., \&
        Ibano{\v g}lu, C.\ 2008, \mnras, 391, 802 
\bibitem[Almeida et al.(2012)]{dvoj59} Almeida, L.~A., 
        Jablonski, F., Tello, J., \& Rodrigues, C.~V.\ 2012, \mnras, 423, 478 
\bibitem[Antokhin et al.(2004)]{igor} Antokhin, I. I., Owocki, S. P., \&
        Brown, J. C., 2004, ApJ, 611, 434
\bibitem[Antokhin et al.(2008)]{igorkar} Antokhin, I. I., Rauw, G.,
         Vreux, J.-M., van der Hucht, K. A., \& Brown, J. C. 2008, A\&A,
         477, 593
\bibitem[Asplund et al.(2009)]{asp09} Asplund, M., Grevesse, N., Sauval, A. J.,
         \& Scott, P., 2009, ARA\&A, 47, 481
\bibitem[Aungwerojwit et al.(2007)]{dvoj29} Aungwerojwit, A., G{\"a}nsicke, B.~T.,
        Rodr{\'{\i}}guez-Gil, P., et al.\ 2007, \aap, 469, 297 
\bibitem[Babel(1996)]{babelb} Babel, J. 1996, A\&A, 309, 867
\bibitem[Barlow et al.(2010)]{dvoj37} Barlow, B.~N., Dunlap, 
        B.~H., Clemens, J.~C., et al.\ 2010, \mnras, 403, 324 
\bibitem[Barlow et al.(2011)]{dvoj38} Barlow, B.~N., Dunlap, 
        B.~H., Clemens, J.~C., et al.\ 2011, \mnras, 414, 3434 
\bibitem[Barlow et al.(2012a)]{dvoj57} Barlow, B.~N., Wade, 
        R.~A., \& Liss, S.~E.\ 2012a, \apj, 753, 101 
\bibitem[Barlow et al.(2012b)]{dvoj39} Barlow, B.~N., Wade, 
        R.~A., Liss, S.~E., {\O}stensen, R.~H., 
        \& Van Winckel, H.\ 2012b, \apj, 758, 58 
\bibitem[Barlow et al.(2013)]{dvoj25} Barlow, B.~N., Kilkenny, 
        D., Drechsel, H., et al.\ 2013, \mnras, 430, 22 
\bibitem[Baschek et al.(1982)]{sam16} Baschek, B., Scholz, M., Kudritzki, R.~P.,
        \& Simon, K.~P.\ 1982, \aap, 108, 387 
\bibitem[Bauer \& Husfeld(1995)]{sam5} Bauer, F., \& Husfeld, D.\ 1995, \aap,
        300, 481 
\bibitem[Bisscheroux et al.(1997)]{dvoj17} Bisscheroux, B.~C., Pols, O.~R.,
        Kahabka, P., Belloni, T., \& van den Heuvel, E.~P.~J.\ 1997, \aap,
        317, 815 
\bibitem[Bloemen et al.(2011)]{dvoj58} Bloemen, S., Marsh, 
        T.~R., {\O}stensen, R.~H., et al.\ 2011, \mnras, 410, 1787 
\bibitem[Blondin et al.(1990)]{blondyn} Blondin, J. M., Kallman, T. R.,
        Fryxell, B. A., \& Taam, R. E. 1990, ApJ, 356, 591
\bibitem[Bouret et al.(2003)]{bourak}  Bouret, J.-C., Lanz, T., Hillier,
        D. J., et al.\ 2003, ApJ, 595, 1182
\bibitem[Budaj et al.(2003)]{dvoj61} Budaj, J., Elkin, V., 
        \& Hubeny, I.\ 2003, Modelling of Stellar Atmospheres, 210, 44P 
\bibitem[Caillault \& Helfand(1985)]{plejadani} Caillault, J.-P., \& Helfand,
        D.~J.\ 1985, \apj, 289, 279 
\bibitem[Castelli(2005)]{casat} Castelli, F. 2005, Memorie della Societ\`a
        Astronomica Italiana Supplement, 8, 25
\bibitem[Castor et al.(1976)]{cak76} Castor, J. I., Abbott, D. C., \&
        Klein, R. I. 1976, Physique des mouvements dans les atmosph\`eres
        stellaires, eds. R.~Cayrel \& M.~Sternberg (Paris: CNRS), 363
\bibitem[{\v C}echura \& Hadrava(2015)]{hacek} {\v C}echura, J., \& Hadrava,
        P.\ 2015, \aap, 575, A5
\bibitem[Charpinet et al.(2002)]{ostrasit} Charpinet, S., 
        Fontaine, G., Brassard, P., \& Dorman, B.\ 2002, \apjs, 140, 469 
\bibitem[Charpinet et al.(2005a)]{sam13} Charpinet, S., Fontaine, G., Brassard,
        P., Green, E.~M., \& Chayer, P.\ 2005a, \aap, 437, 575 
\bibitem[Charpinet et al.(2005b)]{dvoj32} Charpinet, S., Fontaine, G., Brassard,
        P., et al.\ 2005, \aap, 443, 251 
\bibitem[Charpinet et al.(2006)]{sam14} Charpinet, S., Silvotti, R.,
        Bonanno, A., et al.\ 2006, \aap, 459, 565 
\bibitem[Chayer et al.(2015)]{sam31} Chayer, P., Dixon, 
        W.~V., Fullerton, A.~W., Ooghe-Tabanou, B., 
        \& Reid, I.~N.\ 2015, \mnras, 452, 2292 
\bibitem[Cohen et al.(2008)]{cobecru} Cohen D. H., Kuhn M. A., Gagn\'e M.,
        Jensen E. L. N., \& Miller N. A., 2008, MNRAS, 386, 1855
\bibitem[Cooke et al.(1978)]{kapitan} Cooke B. A., Fabian A. C., Pringle J.
        E., 1978, Nature, 273, 645
\bibitem[Cranmer \& Saar(2011)]{crasa} Cranmer, S.~R., \& Saar, S.~H.\ 2011,
        \apj, 741, 54 
\bibitem[Aznar Cuadrado \& Jeffery(2002)]{dvoj51} Aznar Cuadrado, R., \&
        Jeffery, C.~S.\ 2002, \aap, 385, 131 
\bibitem[Daniel et al.(2002)]{daniel} Daniel, K.~J., Linsky, 
        J.~L., \& Gagn{\'e}, M.\ 2002, \apj, 578, 486 
\bibitem[Davidson \& Ostriker(1973)]{davos} Davidson, K., \& Ostriker, J. P.
        1973, ApJ, 179, 585
\bibitem[Deetjen(2000)]{dvoj34} Deetjen, J.~L.\ 2000, \aap, 360, 281 
\bibitem[Donati et al.(2002)]{donati} Donati, J.-F., Babel, 
        J., Harries, T.~J., et al.\ 2002, \mnras, 333, 55 
\bibitem[Dorman et al.(1993)]{durman} Dorman, B., Rood, R.~T., 
        \& O'Connell, R.~W.\ 1993, \apj, 419, 596 
\bibitem[Drake et al.(1991)]{norbert} Drake, S.~A., Linsky, 
        J.~L., Judge, P.~G., \& Elitzur, M.\ 1991, \aj, 101, 230 
\bibitem[Drechsel et al.(2001)]{dvoj20} Drechsel, H., Heber, U.,
        Napiwotzki, R., et al.\ 2001, \aap, 379, 893 
\bibitem[Feldmeier \& Shlosman(2000)]{feslop} Feldmeier, A., \& Shlosman, I.
        2000, ApJL, 532, 125
\bibitem[Feldmeier \& Nikutta(2006)]{felnik} Feldmeier, A., \& Nikutta,
        R.\ 2006, \aap, 446, 661
\bibitem[Feldmeier et al.(1996)]{felabon} Feldmeier, A., Anzer, U., Boerner, G.,
        \& Nagase, F. 1996, A\&A, 311, 793
\bibitem[Feldmeier et al.(1997)]{felpulpal} Feldmeier, A., Puls, J., \&
        Pauldrach, A. W. A. 1997, A\&A, 322, 878
\bibitem[Feldmeier et al.(2008)]{fero} Feldmeier, A., R\"atzel, D., \& Owocki,
        S. P. 2008, ApJ, 679, 704
\bibitem[Fontaine et al.(2014)]{dvoj71} Fontaine, G., Green, 
        E., Charpinet, S., et al.\ 2014, 6th Meeting on Hot Subdwarf Stars and 
        Related Objects, 481, 19 
\bibitem[For et al.(2010)]{dvoj46} For, B.-Q., Green, E.~M., 
        Fontaine, G., et al.\ 2010, \apj, 708, 253 
\bibitem[Fransson \& Fabian(1980)]{ff} Fransson, C., \& Fabian, A.~C. 1980,
        \aap, 87, 102
\bibitem[Friend \& Abbott(1986)]{fa} Friend, D.~B., \& Abbott,
D.~C.\ 1986, \apj, 311, 701 
\bibitem[Geier \& Heber(2012)]{ghrot} Geier, S., \& Heber, U.\ 2012, \aap, 543,
        A149 
\bibitem[Geier et al.(2007)]{dvoj55} Geier, S., Nesslinger, S., Heber, U.,
        et al.\ 2007, \aap, 464, 299 
\bibitem[Geier et al.(2008)]{dvoj1} Geier, S., Nesslinger, S., Heber, U., et al.\
        2008, \aap, 477, L13 
\bibitem[Geier et al.(2009)]{sam17} Geier, S., Edelmann, H., 
        Heber, U., \& Morales-Rueda, L.\ 2009, \apjl, 702, L96 
\bibitem[Geier et al.(2010a)]{dvoj3} Geier, S., Heber, U., Kupfer, T., \& Napiwotzki,
        R.\ 2010, \aap, 515, A37 
\bibitem[Geier et al.(2010b)]{dvoj65} Geier, S., Heber, U., Podsiadlowski, P.,
        et al.\ 2010b, \aap, 519, A25 
\bibitem[Geier et al.(2014)]{dvoj12} Geier, S., {\O}stensen, R.~H., Heber, U., et
        al.\ 2014, \aap, 562, A95 
\bibitem[Geier et al.(2015)]{dvoj23} Geier, S., F{\"u}rst, F., 
        Ziegerer, E., et al.\ 2015, Science, 347, 1126 
\bibitem[Gies et al.(1998)]{dvoj4} Gies, D.~R., Bagnuolo, 
        W.~G., Jr., Ferrara, E.~C., et al.\ 1998, \apj, 493, 440 
\bibitem[Granada et al.(2013)]{granada} Granada, A., Ekstr\"om, S., Georgy, C.,
        et al. 2013, A\&A, 553, A25
\bibitem[Green et al.(1984)]{dvoj64} Green, R.~F., Liebert, 
        J., \& Wesemael, F.\ 1984, \apj, 280, 177 
\bibitem[Gruschinske et al.(1983)]{dvoj43} Gruschinske, J., Hamann, W.~R.,
        Kudritzki, R.~P., Simon, K.~P., \& Kaufmann, J.~P.\ 1983, \aap, 121, 85 
\bibitem[Hamann(2010)]{dvoj69} Hamann, W.-R.\ 2010, \apss, 329, 151
\bibitem[Hamann et al.(1981)]{sam32} Hamann, W.-R., Gruschinske, J., Kudritzki,
        R.~P., \& Simon, K.~P.\ 1981, \aap, 104, 249 
\bibitem[Han et al.(2007)]{jinanvoxfordu} Han, Z., Podsiadlowski, P., 
        \& Lynas-Gray, A.~E.\ 2007, \mnras, 380, 1098 
\bibitem[Harmanec(1988)]{har} Harmanec, P. 1988, Bull. Astron. Inst. Czechosl.
        39, 329
\bibitem[Heber et al.(1999)]{dvoj47} Heber, U., Reid, I.~N., \& Werner,
        K.\ 1999, \aap, 348, L25 
\bibitem[Heber et al.(2000)]{sam22} Heber, U., Reid, I.~N., \& Werner, K.\ 2000,
        \aap, 363, 198 
\bibitem[Heber et al.(2004)]{dvoj63} Heber, U., Drechsel, H., {\O}stensen, R.,
        et al.\ 2004, \aap, 420, 251 
\bibitem[Hilditch et al.(1996)]{dvoj28} Hilditch, R.~W., 
        Harries, T.~J., \& Hill, G.\ 1996, \mnras, 279, 1380 
\bibitem[Hirsch(2009)]{sam6} Hirsch H., 2009, PhD thesis, University of
        Erlangen-N\"urnberg
\bibitem[Holzwarth \& Jardine(2007)]{nemskot} Holzwarth, V., \& Jardine,
        M.\ 2007, \aap, 463, 11 
\bibitem[Howarth \& Heber(1990)]{dvoj44} Howarth, I.~D., \& Heber, U.\ 1990,
        \pasp, 102, 912 
\bibitem[Hummer et al.(1993)]{zel0} Hummer, D. G., Berrington, K. A.,
        Eissner, W. et al. 1993, A\&A, 279, 298
\bibitem[Husfeld et al.(1989)]{sam15} Husfeld, D., Butler, K., Heber, U.,
        \& Drilling, J.~S.\ 1989, \aap, 222, 150 
\bibitem[{\.I}bano{\v g}lu et al.(2004)]{dvoj35} {\.I}bano{\v g}lu, C.,
        {\c C}ak{\i}rl{\i}, {\"O}., Ta{\c s}, G., \& Evren, S.\ 2004, \aap, 414, 1043 
\bibitem[Iben et al.(1983)]{icko} Iben, I., Jr., Kaler, 
        J.~B., Truran, J.~W., \& Renzini, A.\ 1983, \apj, 264, 605 
\bibitem[Iben \& Tutukov(1984)]{iba} Iben, I., Jr., \& Tutukov, A.~V.\ 1984,
        \apjs, 54, 335 
\bibitem[Jeffery \& Hamann(2010)]{jefham} Jeffery, C.~S., \& Hamann, W.-R.\
        2010, \mnras, 404, 1698 
\bibitem[Jeffery et al.(2015)]{sam21} Jeffery, C.~S., Ahmad, 
        A., Naslim, N., \& Kerzendorf, W.\ 2015, \mnras, 446, 1889 
\bibitem[Koen et al.(1998)]{dvoj10} Koen, C., Orosz, J.~A., 
        \& Wade, R.~A.\ 1998, \mnras, 300, 695 
\bibitem[Koubsk{\'y} et al.(2012)]{koukov} Koubsk{\'y}, P., Kotkov{\'a}, L.,
        Votruba, V., {\v S}lechta, M., \& Dvo{\v r}{\'a}kov{\'a},
        {\v S}.\ 2012, \aap, 545, A121 
\bibitem[Koubsk{\'y} et al.(2014)]{dalsikkk} Koubsk{\'y}, P., Kotkov{\'a}, L.,
        Kraus, M., et al.\ 2014, \aap, 567, A57 
\bibitem[Krti\v cka(2006)]{nlteii} Krti\v cka, J. 2006, MNRAS, 367, 1282
\bibitem[Krti{\v c}ka(2014)]{metuje} Krti{\v c}ka, J.\ 2014, \aap, 564, A70
\bibitem[Krti\v cka \& Kub\'at(2009)]{nlteiii} Krti\v cka, J., \& Kub\'at,
        J. 2009, MNRAS, 394, 2065
\bibitem[Krti\v cka \& Kub\'at(2010a)]{cmf1} Krti\v cka, J., \&  Kub\'at,
        J. 2010a, A\&A, 519, A5
\bibitem[Krti{\v c}ka \& Kub{\'a}t(2010b)]{krtzatmeni} Krti{\v c}ka, J.,
        \& Kub{\'a}t, J.\ 2010b, \apss, 329, 145 
\bibitem[Krti{\v c}ka \& Kub{\'a}t(2011)]{betyna} Krti{\v c}ka, J.,
        \& Kub{\'a}t, J.\ 2011, \aap, 534, A97
\bibitem[Krti\v{c}ka et al.(2009)]{simx} Krti\v{c}ka J., Feldmeier A.,
        Oskinova L. M., Kub\'at J., Hamann W.-R., 2009, A\&A, 508, 841
\bibitem[Krti\v cka et al.(2011)]{kom} Krti\v{c}ka,~J., Owocki, S. P., \&
       Meynet, G., 2011, A\&A, 527, A84
\bibitem[Krti{\v c}ka et al.(2015)]{dvojvit} Krti{\v c}ka, J., Kub{\'a}t, J.,
        \& Krti{\v c}kov{\'a}, I.\ 2015, \aap, 579, A111 
\bibitem[Kub\'at(2003)]{kub} Kub\'at J., 2003, in Piskunov N. E., Weiss W. W.,
        Gray D. F., Modelling of Stellar Atmospheres, IAU Symp. 210. ASP,
        San Francisco, p.~A8
\bibitem[Kub\'at et al.(1999)]{kpp} Kub\'at, J., Puls, J., \& Pauldrach,
       A. W. A. 1999, A\&A, 341, 587
\bibitem[Kudritzki \& Simon(1978)]{dvoj16} Kudritzki, R.~P., \& Simon, K.~P.\ 1978,
        \aap, 70, 653 
\bibitem[Kupfer et al.(2015)]{dvoj49} Kupfer, T., Geier, S., Heber, U., et
        al.\ 2015, \aap, 576, A44 
\bibitem[Kupka et al.(1999)]{vald2} Kupka, F., Piskunov, N. E., Ryabchikova,
        T. A., Stempels, H. C., \& Weiss, W. W. 1999, A\&AS, 138, 119
\bibitem[Kurf{\"u}rst et al.(2014)]{kurfek} Kurf{\"u}rst, P.,
        Feldmeier, A., \& Krti{\v c}ka, J.\ 2014, \aap, 569, A23
\bibitem[Kurucz(2005)]{kurat} Kurucz, R. L. 2005, Memorie della Societ\`a
        Astronomica Italiana Supplement, 8, 1
\bibitem[Landstreet \& Borra(1978)]{labor} Landstreet, J. D. \& Borra, E. F.
    1978, ApJL, 224, 5
\bibitem[Landstreet et al.(1998)]{nedolez} Landstreet, J.~D., Dolez, N., \&
        Vauclair, S. 1998, A\&A, 333, 977
\bibitem[Landstreet et al.(2012)]{podland} Landstreet, J.~D., Bagnulo, S.,
        Fossati, L., Jordan, S., \& O'Toole, S.~J.\ 2012, \aap, 541, A100 
\bibitem[La Palombara et al.(2012)]{dvoj5} La Palombara, N., 
        Mereghetti, S., Tiengo, A., \& Esposito, P.\ 2012, \apjl, 750, L34 
\bibitem[La Palombara et al.(2014)]{bufacek} La Palombara, N., Esposito, P.,
       Mereghetti, S., \& Tiengo, A.\ 2014, \aap, 566, A4 
\bibitem[La Palombara et al.(2015)]{sam11} La Palombara, N., Esposito, P.,
        Mereghetti, S., Novara, G., \& Tiengo, A.\ 2015, \aap, 580, A56 
\bibitem[Lamers et al.(1976)]{laheupet} Lamers, H. J. G. L. M., van den Heuvel,
        E. P. J., Petterson, J. A. 1976, A\&A, 49, 327
\bibitem[Lamers et al.(1995)Lamers, Snow, \& Lindholm]{lsl} Lamers,
        H. J. G. L. M., Snow, T. P., \& Lindholm, D. M. 1995, ApJ, 455, 269
\bibitem[Lanz et al.(1997)]{sam8} Lanz, T., Hubeny, I., 
        \& Heap, S.~R.\ 1997, \apj, 485, 843 
\bibitem[Lanz \& Hubeny(2003)]{ostar2003} Lanz, T.,  \& Hubeny, I. 2003, ApJS,
        146, 417
\bibitem[Lanz et al.(2004)]{sam4} Lanz, T., Brown, T.~M., 
        Sweigart, A.~V., Hubeny, I., \& Landsman, W.~B.\ 2004, \apj, 602, 342 
\bibitem[Lanz \& Hubeny(2007)]{bstar2006} Lanz, T.,  \& Hubeny, I. 2007,
        ApJS, 169, 83
\bibitem[Latour et al.(2013)]{sam25} Latour, M., Fontaine, 
        G., Chayer, P., \& Brassard, P.\ 2013, \apj, 773, 84 
\bibitem[Latour et al.(2014)]{dvoj33} Latour, M., Fontaine, 
        G., Green, E.~M., Brassard, P., \& Chayer, P.\ 2014, \apj, 788, 65 
\bibitem[Lucy(2012)]{lucyjakomy} Lucy L. B. 2012, A\&A, 544, A120
\bibitem[Lucy \& Solomon(1970)]{lusol} Lucy, L. B., \& Solomon, P. M. 1970,
        ApJ, 159, 879
\bibitem[Lucy \& White(1980)]{luciebila} Lucy L. B., \& White, R. L.,
        1980, ApJ, 241, 300
\bibitem[Madura \& Groh(2012)]{magro} Madura, T.~I., \& Groh, J.~H.\ 2012,
        \apjl, 746, L18 
\bibitem[Martins et al.(2005)]{martclump} Martins, F., Schaerer, D.,
        Hillier, D. J., et al. 2005, A\&A, 441, 735
        Tsymbal, V., \& Mkrtichian, D. E. 2007, A\&A, 471, 941
\bibitem[Maxted et al.(2000)]{dvoj54} Maxted, P.~F.~L., Marsh, 
        T.~R., \& North, R.~C.\ 2000, \mnras, 317, L41 
\bibitem[Maxted et al.(2002)]{dvoj68} Maxted, P.~F.~L., Marsh, 
        T.~R., Heber, U., et al.\ 2002, \mnras, 333, 231 
\bibitem[Mereghetti et al.(2011)]{dvoj66} Mereghetti, S., Campana, S.,
        Esposito, P., La Palombara, N., \& Tiengo, A.\ 2011, \aap, 536, A69 
\bibitem[Mereghetti et al.(2013)]{dvoj18} Mereghetti, S., La Palombara, N.,
        Tiengo, A., et al.\ 2013, \aap, 553, A46 
\bibitem[Mereghetti et al.(2014)]{merne} Mereghetti, S., La 
        Palombara, N., Esposito, P., et al.\ 2014, \mnras, 441, 2684 
\bibitem[Mereghetti et al.(2016)]{dvoj72} Mereghetti, S., Pintore, F.,
        Esposito, P. et al. 2016, submitted to \mnras
\bibitem[Michaud et al.(2011)]{miriri} Michaud, G., Richer, J., \& Richard,
        O.\ 2011, \aap, 529, A60 
\bibitem[Mihalas \& Mihalas(1999)]{mihmih} Mihalas, D., Mihalas, B., 1999,
        Foundations of Radiation Hydrodynamics (New York: Dower)
\bibitem[Mihalas et al.(1975)]{mikuh} Mihalas, D., Kunasz, P. B., \&
        Hummer, D. G. 1975, ApJ, 202, 465
\bibitem[Miller Bertolami \& Althaus(2006)]{argentinci} Miller Bertolami, M.~M.,
        \& Althaus, L.~G.\ 2006, \aap, 454, 845 
\bibitem[Montez et al.(2010)]{dvoj9} Montez, R., Jr., De 
        Marco, O., Kastner, J.~H., \& Chu, Y.-H.\ 2010, \apj, 721, 1820 
\bibitem[Napiwotzki(1999)]{dvoj2} Napiwotzki, R.\ 1999, \aap, 350, 101 
\bibitem[Naslim et al.(2010)]{sam27} Naslim, N., Jeffery, 
        C.~S., Ahmad, A., Behara, N.~T., 
\& {\c S}ah{\`i}n, T.\ 2010, \mnras, 409, 582
\bibitem[Naslim et al.(2011)]{sam20} Naslim, N., Jeffery, 
        C.~S., Behara, N.~T., \& Hibbert, A.\ 2011, \mnras, 412, 363 
\bibitem[Naz{\'e}(2009)]{naze} Naz{\'e}, Y.\ 2009, \aap, 506, 1055 
\bibitem[Naz{\'e} et al.(2012)]{nazcyg} Naz{\'e}, Y., Mahy, L., Damerdji, Y.,
        et al.\ 2012, \aap, 546, A37 
\bibitem[Naz{\'e} et al.(2014)]{malanaze} Naz{\'e}, Y., Petit, 
        V., Rinbrand, M., et al.\ 2014, \apjs, 215, 10 
\bibitem[Okazaki(2001)]{sapporo} Okazaki, A. T. 2001, PASJ, 53, 119
\bibitem[Orosz \& Wade(1999)]{dvoj11} Orosz, J.~A., \& Wade, R.~A.\ 1999, \mnras,
        310, 773 
\bibitem[{\O}stensen(2006)]{sam2} {\O}stensen, R.~H.\ 2006, 
        Baltic Astronomy, 15, 85 
\bibitem[{\O}stensen et al.(2008)]{dvoj62} {\O}stensen, R.~H., 
        Oreiro, R., Hu, H., Drechsel, H., 
        \& Heber, U.\ 2008, Hot Subdwarf Stars and Related Objects, 392, 221 
\bibitem[{\O}stensen et al.(2010)]{dvoj56} {\O}stensen, R.~H., 
        Green, E.~M., Bloemen, S., et al.\ 2010, \mnras, 408, L51 
\bibitem[Owocki \& Cohen(1999)]{oskal} Owocki, S. P., \& Cohen, D. H. 1999, ApJ,
        520, 833
\bibitem[Owocki \& Puls(2002)]{op} Owocki, S. P., \& Puls, J., 2002, ApJ, 568,
        965
\bibitem[Owocki \& Rybicki(1984)]{owor} Owocki, S.~P., \& Rybicki, G.~B.\ 1984,
        \apj, 284, 33
\bibitem[Owocki et al.(1988)]{ocr} Owocki, S. P., Castor, J. I., \&
        Rybicki, G. B. 1988, ApJ, 335, 914
\bibitem[Owocki et al.(2013)]{owomix} Owocki, S.~P., Sundqvist, J.~O., Cohen,
        D.~H., \& Gayley, K.~G.\ 2013, \mnras, 429, 3379 
\bibitem[Parkin \& Sim(2013)]{pars} Parkin, E.~R., \& Sim, S.~A.\ 2013, \apj,
        767, 114
\bibitem[Parkin et al.(2014)]{papina} Parkin, E.~R., Pittard, J.~M.,
        Naz{\'e}, Y., \& Blomme, R.\ 2014, \aap, 570, A10 
\bibitem[Pauldrach \& Puls(1990)]{bista} Pauldrach, A. W. A., \& Puls,
        J. \citeyear{bista}, A\&A 237, 409
\bibitem[Pauldrach et al.(1986)]{ppk} Pauldrach, A., Puls, J., \& Kudritzki, R.
       P. 1986, A\&A, 164, 86
\bibitem[Pauldrach et al.(2001)]{pahole} Pauldrach, A. W. A., Hoffmann, T. L.,
        \& Lennon M. 2001, A\&A, 375, 161
\bibitem[Pauldrach et al.(2004)]{btpau} Pauldrach, A. W. A., Hoffmann, T. L., \&
        M\'endez, R. H. 2004, A\&A, 419, 1111
\bibitem[Paxton et al.(2011)]{mesa1} Paxton, B., Bildsten,
        L., Dotter, A., et al.\ 2011, \apjs, 192, 3
\bibitem[Paxton et al.(2013)]{mesa2} Paxton, B., Cantiello,
        M., Arras, P., et al.\ 2013, \apjs, 208, 4
\bibitem[Peters et al.(2008)]{dvoj21} Peters, G.~J., Gies, 
        D.~R., Grundstrom, E.~D., \& McSwain, M.~V.\ 2008, \apj, 686, 1280 
\bibitem[Peters et al.(2013)]{dvoj60} Peters, G.~J., Pewett, 
        T.~D., Gies, D.~R., Touhami, Y.~N., \& Grundstrom, E.~D.\ 2013, \apj,
        765, 2 
\bibitem[Petit et al.(2013)]{malykor} Petit, V., Owocki, S. P., Wade, G. A. et
        al. 2013, MNRAS, 429, 398
\bibitem[Piskunov et al.(1995)]{vald1} Piskunov, N. E., Kupka, F., Ryabchikova,
        T. A., Weiss, W. W., \& Jeffery, C. S. 1995, A\&AS, 112, 525
\bibitem[Pittard(2009)]{skoda} Pittard, J. M. 2009, MNRAS, 396, 1743
\bibitem[Prilutskii \& Usov(1976)]{usaci} Prilutskii O. F., \& Usov V. V.,
        1976, AZh, 53, 6
\bibitem[Przybilla et al.(2005)]{sam30} Przybilla, N., Butler, K., Heber, U., \&
        Jeffery, C.~S.\ 2005, \aap, 443, L25 
\bibitem[Puls et al.(2000)]{pusle} Puls, J., Springmann, U., \& Lennon, M. 2000,
        A\&AS, 141, 23
\bibitem[Puls et al.(2008)]{pulvina} Puls, J., Vink, J.~S., \& Najarro, F.\ 2008,
        \aapr, 16, 209 
\bibitem[Quirrenbach et al.(1997)]{quir} Quirrenbach, A., 
        Bjorkman, K.~S., Bjorkman, J.~E., et al.\ 1997, \apj, 479, 477 
\bibitem[Randall et al.(2007)]{sam9} Randall, S.~K., Green, E.~M., Van Grootel,
        V., et al.\ 2007, \aap, 476, 1317 
\bibitem[Randall et al.(2009)]{sam12} Randall, S.~K., Van Grootel, V.,
        Fontaine, G., Charpinet, S., \& Brassard, P.\ 2009, \aap, 507, 911 
\bibitem[Rauch(1993)]{dvoj45} Rauch, T.\ 1993, \aap, 276, 171 
\bibitem[Rauch(2000)]{dvoj13} Rauch, T.\ 2000, \aap, 356, 665 
\bibitem[Rauch \& Werner(2003)]{dvoj14} Rauch, T., \& Werner, K.\ 2003, \aap, 400,
        271 
\bibitem[Rauch et al.(2014)]{sam26} Rauch, T., Rudkowski, A., Kampka, D.,
        et al.\ 2014, \aap, 566, A3 
\bibitem[Raymond \& Smith(1977)]{rs} Raymond J. C., \& Smith B. W., 1977, ApJS,
        35, 419
\bibitem[Ribeiro \& Baptista(2011)]{dvoj30} Ribeiro, T., \& Baptista, R.\ 2011,
        \aap, 526, A150 
\bibitem[Rivinius et al.(2013)]{ricam} Rivinius, T., Carciofi, A.~C., \&
        Martayan, C.\ 2013, \aapr, 21, 69
\bibitem[Rodr{\'{\i}}guez-L{\'o}pez et al.(2007)]{sam18} 
        Rodr{\'{\i}}guez-L{\'o}pez, C., Ulla, A., 
        \& Garrido, R.\ 2007, \mnras, 379, 1123 
\bibitem[Rybicki \& Hummer(1978)]{rybashumrem} Rybicki, G. B., \& Hummer, D. G.,
        1978, ApJ, 219, 645
\bibitem[Saffer et al.(1994)]{dvoj67} Saffer, R.~A., Bergeron, 
        P., Koester, D., \& Liebert, J.\ 1994, \apj, 432, 351 
\bibitem[Saio \& Jeffery(2000)]{saje} Saio, H., \& Jeffery, C.~S.\ 2000,
        \mnras, 313, 671 
\bibitem[Schaffenroth et al.(2011)]{dvoj22} Schaffenroth, V., 
        Geier, S., Heber, U., et al.\ 2011, American Institute of Physics 
        Conference Series, 1331, 174 
\bibitem[Schaffenroth et al.(2013)]{dvoj24} Schaffenroth, V., Geier, S.,
        Drechsel, H., et al.\ 2013, \aap, 553, A18 
\bibitem[Schaffenroth et al.(2014)]{dvoj48} Schaffenroth, V., Geier, S., Heber,
        U., et al.\ 2014, \aap, 564, A98 
\bibitem[Schaffenroth et al.(2015)]{dvoj36} Schaffenroth, V., Barlow, B.~N.,
        Drechsel, H., \& Dunlap, B.~H.\ 2015, \aap, 576, A123 
\bibitem[Schindler et al.(2015)]{svitavy} Schindler, J.-T., 
        Green, E.~M., \& Arnett, W.~D.\ 2015, \apj, 806, 178 
\bibitem[Schmitt \& Liefke(2004)]{humburk} Schmitt, J.~H.~M.~M.,
        \& Liefke, C.\ 2004, \aap, 417, 651 
\bibitem[Schonberner \& Drilling(1984)]{sam29} Schonberner, D., \&
        Drilling, J.~S.\ 1984, \apj, 278, 702 
\bibitem[Schure et al.(2009)]{jiste} Schure, K.~M., Kosenko, D., Kaastra, J.~S.,
        Keppens, R., \& Vink, J.\ 2009, \aap, 508, 751 
\bibitem[Shimanskii et al.(2004)]{dvoj8} Shimanskii, V.~V., Borisov, N.~V.,
        Sakhibullin, N.~A., \& Surkov, A.~E.\ 2004, Astronomy Reports, 48, 563 
\bibitem[Seaton et al.(1992)]{topt} Seaton, M. J., Zeippen, C. J., Tully, J. A.
        et al. 1992, Rev. Mexicana Astron. Astrofis., 23, 19
\bibitem[Springmann \& Pauldrach(1992)]{treni} Springmann, U. W. E.,
        \& Pauldrach, A. W. A., 1992, A\&A 262, 515
\bibitem[Telting et al.(2012)]{dvoj53} Telting, J.~H., {\O}stensen, R.~H.,
        Baran, A.~S., et al.\ 2012, \aap, 544, A1 
\bibitem[Theissen et al.(1995)]{dvoj50} Theissen, A., Moehler, S., Heber, U.,
        Schmidt, J.~H.~K., \& de Boer, K.~S.\ 1995, \aap, 298, 577 
\bibitem[Thejll et al.(1994)]{sam1} Thejll, P., Bauer, F., 
        Saffer, R., et al.\ 1994, \apj, 433, 819 
\bibitem[Thejll et al.(1995)]{dvoj26} Thejll, P., Ulla, A.,
        \& MacDonald, J.\ 1995, \aap, 303, 773 
\bibitem[Townsend et al.(2005)]{towog} Townsend, R.\,H.\,D. Owocki, S.\,P., \&
        Groote D. 2005, ApJ, 630, L81
\bibitem[ud-Doula \& Owocki(2002)]{udo} ud-Doula, A., \& Owocki, S.~P.\ 2002,
        \apj, 576, 413 
\bibitem[ud-Doula et al.(2008)]{udorot} ud-Doula, A., Owocki, 
        S.~P., \& Townsend, R.~H.~D.\ 2008, \mnras, 385, 97 
\bibitem[ud-Doula et al.(2009)]{brzdud} ud-Doula, A., Owocki, S. P.,
        \& Townsend, R. H. D. 2009, MNRAS, 392, 1022
\bibitem[Ulla \& Thejll(1998)]{sam19} Ulla, A., \& Thejll, P.\ 1998, \aaps, 132,
        1 
\bibitem[Unglaub(2008)]{un} Unglaub, K.\ 2008, \aap, 486, 923
\bibitem[Vauclair(1975)]{vasam} Vauclair, S. 1975, A\&A, 45, 233
\bibitem[Van Grootel et al.(2010)]{sam28} Van Grootel, V., 
        Charpinet, S., Fontaine, G., et al.\ 2010, \apjl, 718, L97 
\bibitem[Van Grootel et al.(2013)]{dvoj41} Van Grootel, V., Charpinet, S.,
        Brassard, P., Fontaine, G., \& Green, E.~M.\ 2013, \aap, 553, A97 
\bibitem[Vennes et al.(2007)]{dvoj15} Vennes, S., Kawka, A., 
        \& Smith, J.~A.\ 2007, \apjl, 668, L59 
\bibitem[Vick et al.(2011)]{vimiri} Vick, M., Michaud, G., Richer, J., \&
        Richard, O. 2011, A\&A, 526, A37
\bibitem[Vink \& Cassisi(2002)]{vinca} Vink, J.~S., \& Cassisi, S.\ 2002, \aap,
        392, 553 
\bibitem[Vink et al.(1999)]{vikolabis} Vink, J.~S., de Koter, A., \& Lamers,
        H.~J.~G.~L.~M.\ 1999, \aap, 350, 181 
\bibitem[Vink et al.(2001)]{vikolamet} Vink, J. S., de Koter, A., \& Lamers,
          H. J. G. L. M. 2001, A\&A, 369, 574
\bibitem[Viton et al.(1988)]{dvoj7} Viton, M., Burgarella, D., Cassatella, A., \&
        Prevot, L.\ 1988, \aap, 205, 147 
\bibitem[Vos et al.(2012)]{dvoj31} Vos, J., {\O}stensen, R.~H., Degroote, P., et
        al.\ 2012, \aap, 548, A6 
\bibitem[Vos et al.(2013)]{dvoj19} Vos, J., {\O}stensen, R.~H., N{\'e}meth, P., et
        al.\ 2013, \aap, 559, A54 
\bibitem[Votruba et al.(2007)]{ufo} Votruba, V., Feldmeier, A., Kub\'at, J.,
        \& R\"{a}tzel, D. 2007, A\&A, 474, 549
\bibitem[Votruba et al.(2010)]{votzameni} Votruba, V., Feldmeier, A.,
        Krti{\v c}ka, J., \& Kub{\'a}t, J.\ 2010, \apss, 329, 159 
\bibitem[Vu{\v c}kovi{\'c} et al.(2007)]{dvoj40} Vu{\v c}kovi{\'c}, M., Aerts,
        C., {\"O}stensen, R., et al.\ 2007, \aap, 471, 605 
\bibitem[Vu{\v c}kovi{\'c} et al.(2015)]{vuckodor} Vu{\v 
        c}kovi{\'c}, M., {\O}stensen, R.~H., N{\'e}meth, P., Bloemen, S., 
        \& P{\'a}pics, P.~I.\ 2015, A\&A, in press
\bibitem[Wood et al.(2005)]{strelec} Wood, B.~E., M{\"u}ller, 
        H.-R., Zank, G.~P., Linsky, J.~L., \& Redfield, S.\ 2005, \apjl, 628,
        L143 
\bibitem[Zel'dovich \& Raizer(2002)]{zelda} Zel'dovich, Y. B., Raizer, Y. P.
        2002, Physics of Shock Waves and High-temperature Hydrodynamic Phenomena
        (Dover: New York)
\bibitem[Zhang \& Jeffery(2012)]{zhaff} Zhang, X., \& Jeffery, C.~S.\ 2012,
        \mnras, 419, 452 
\bibitem[Zhekov \& Skinner(2000)]{zhek} Zhekov, S.~A., \& Skinner, S.~L.\ 2000,
        \apj, 538, 808 
\bibitem[Zhou et al.(2006)]{sam24} Zhou, A.-Y., Reed, M.~D., 
        Harms, S., et al.\ 2006, \mnras, 367, 179 
\bibitem[Zhu et al.(2015)]{zhumag} Zhu, C., Pakmor, R., van 
        Kerkwijk, M.~H., \& Chang, P.\ 2015, \apjl, 806, L1 
\end{thebibliography}
